\documentclass[final,authoryear,3p,times]{elsarticle}

\usepackage{amssymb}

\usepackage{graphicx}
\usepackage{subfig}
\usepackage{amsmath,amsthm}
\usepackage{enumitem}

\journal{Insurance, Mathematics and Economics}

\newtheorem{theorem}{Theorem}[section]
\newtheorem{corollary}[theorem]{Corollary}
\numberwithin{equation}{section}

\def \noi {\noindent}

\def \P {{\bf P}}
\def \E {{\bf E}}
\def \pv {\pi^*}
\def \la {\lambda}

\def \eps {\epsilon}
\def \del {\delta}
\def \gam {\gamma}
\def \hp {\hat \psi}
\def \hpz {\hat \psi_0}
\def \hpo {\hat \psi_1}
\def \hpt {\hat \psi_2}
\def \Bone {B^{(1)}}
\def \Btwo {B^{(2)}}
\def \Bthr {B^{(3)}}
\def \R {{\bf R}}
\def \L {{\cal L}}
\def \M {{\cal M}}
\def \D {{\cal D}}

\def \dd {{ \mathrm{d}}}

\def\sqr#1#2{{\vcenter{\vbox{\hrule height.#2pt\hbox{\vrule width.#2pt height#1pt \kern#1pt\vrule width.#2pt}\hrule height.#2pt}}}}
\def \square{\hfill\mathchoice\sqr56\sqr56\sqr{4.1}5\sqr{3.5}5}
\def \qed {$\square$ \medskip}
\def \pf {\noindent  {\it Proof}.\quad }

\def \rem#1 {\medskip \noi {\bf Remark #1.  }}

\begin{document}

\begin{frontmatter}



\title{Minimizing the Probability of Lifetime Ruin under Stochastic Volatility \tnoteref{lab:thanks}}
\tnotetext[lab:thanks]{We thank The Actuarial Foundation for financially supporting this research. We also thank the referee for insightful comments which helped us improve our paper.} 
\author{Erhan Bayraktar\corref{cor1}, Xueying Hu, Virginia R. Young } 
\ead{\{erhan, xyhu, vryoung\}@umich.edu}
\address{Department of Mathematics, University of Michigan, 530 Church Street, Ann Arbor, MI 48104, USA}






\date{}

\begin{abstract}
We assume that an individual invests in a financial market with one riskless and one risky asset, with the latter's price following a diffusion with stochastic volatility. Given the rate of consumption, we  find the optimal investment strategy for the individual who wishes to minimize the probability of going bankrupt.  To solve this minimization problem, we use techniques from stochastic optimal control.
\end{abstract}


\end{frontmatter}


\section{ Introduction}

Pension actuaries traditionally have computed the liabilities for defined benefit (DB) pension plans; however, more and more employees are participating in defined contribution (DC) plans.  Indeed, in June 2007, the Employee Benefits Research Institute (EBRI) reported that in 1979, among active workers participating in retirement plans, the percentages in DB plans only, DC plans only, and both DB and DC plans were 62\%, 16\%, and 22\%, respectively.  The corresponding percentages in 2005 were 10\%, 63\%, and 27\%, respectively.

In terms of numbers of employees, EBRI reported that in 1980, 30.1 million active workers participated in DB plans, while 18.9 million workers participated in DC plans.  The corresponding numbers in 2004 were 20.6 and 52.2 active workers, respectively.  Finally, in terms of numbers of plans in the private sector, in 1980, there were 148 thousand DB plans and 341 thousand DC plans; the corresponding numbers in 2004 were 47 and 653 thousand plans, respectively.

Therefore, however one measures the change in employee coverage under DB versus DC plans, it is clear that pension actuaries will need to adapt to the migration from DB to DC plans.  One way that they can adapt is to switch from advising employers about their DB liabilities to providing investment advice for retirees and employees in DC plans.  The purpose of our proposed research is to help train actuaries for this opportunity under the easy-to-explain goal of an employee or retiree avoiding bankruptcy.

Previous work focused on finding the optimal investment strategy to minimize the probability of bankruptcy under a variety of situations:  (1) allowing the individual to invest in a standard Black-Scholes financial market with a rate of consumption given by some function of wealth, \citep{young,MR2295829}; (2) incorporating immediate and deferred annuities in the financial market, \citep{MR2253122, Bayraktar_Young-naaj09}; (3) limiting borrowing or requiring that borrowing occur at a higher rate than lending, \citep{MR2324574}; (4) modeling consumption as an increasing function of wealth or as a random process correlated with the price process of the stock, \citep{Bayraktar_Young-naaj08, Bayraktar_Moore_Young, Bayraktar_Young-fs09}.  Throughout this body  of work,  the price process of the stock is modeled as a geometric Brownian motion, which is arguably unrealistic, but has given results that one can consider to be ``first approximations.'' Here we extend some of the previous work and allow the stock price to exhibit stochastic volatility. Additionally, we intend to find easy-to-implement rules that will result in nearly minimal probabilities of bankruptcy under stochastic volatility.

The rest of the paper is organized as follows. In Section 2, we introduce the financial market and define the problem of minimizing the probability of lifetime ruin.  In Section 3, we present a related optimal controller-stopper problem, and show that the solution of that problem is the Legendre dual of the minimum probability of lifetime ruin.  
By solving the optimal controller-stopper problem, we effectively solve the problem of minimizing the probability of lifetime ruin. Relying on the results in Section 3, we find an asymptotic approximation of the minimum probability of ruin and the optimal strategy in Section 4. On the other hand, in Section 5, relying on the Markov Chain Approximation Method, we construct a numerical algorithm that solves the original optimal control problem numerically. In Section 6, we present some numerical experiments.

We learn that the optimal investment strategy in the presence of stochastic volatility is not necessarily to invest less in the risky asset than when volatility is fixed. We also observe that the minimal probability of ruin can be almost attained by the asymptotic approximation described in Section 5.1. Also, if an individual uses the investment prescribed by the optimal investment strategy for the constant volatility environment while updating the volatility variable in this formula according to her observations, it turns out she can almost achieve the minimum probability of ruin in a stochastic volatility environment.

\section{The Financial Market and the Probability of Lifetime Ruin}

In this section, we present the financial ingredients that make up the individual's wealth, namely, consumption, a riskless asset, and a risky asset. We, then, define the minimum probability of lifetime ruin.

We assume that the individual invests in a riskless asset whose price at time $t$, $X_t$, follows the process $dX_t = rX_t dt, X_0 = x > 0$, for some fixed rate of interest $r > 0$.  Also, the individual invests in a risky asset whose price at time $t$, $S_t$, follows a diffusion given by
\begin{equation}
\dd S_t =  S_t \left( \mu dt + \sigma_t \, \dd \Bone_t \right), \quad S_0 = S > 0,
\label{eqn:st}
\end{equation}
in which $\mu > r$ and $\sigma_t$ is the (random) volatility of the price process at time $t$.  Here, $\Bone$ is a standard Brownian motion with respect to a filtered probability space $(\Omega, {\cal F}, \P, {\bf F} = \{{\cal F}_t\}_{t \ge 0})$.  We assume that the stochastic volatility is given by
\begin{equation}
\sigma_t = f(Y_t, Z_t),
\label{eqn:sigmat}
\end{equation}
in which $f$ is a smooth positive function that is bounded and bounded away from zero, and $Y$ and $Z$ are two diffusions.  Below, we follow \citet{Fouque:asymp} in specifying the dynamics of $Y$ and $Z$.  Note that if $f$ is constant, then $S$ follows geometric Brownian motion, and that case is considered by \citet{young}.

The first diffusion $Y$ is a fast mean-reverting Gaussian Ornstein-Uhlenbeck process.  Denote by $1/\eps$ the rate of mean reversion of this process, with $0 < \eps \ll 1$ corresponding to the time scale of the process.  $Y$ is an ergodic process, and we assume that its invariant distribution is independent of $\eps$.  In particular, the invariant distribution is normal with mean $m$ and variance $\nu^2$.  The resulting dynamics of $Y$ are given by
\begin{equation}
\dd Y_t = \frac{1}{\eps} \left( m - Y_t \right) dt + \nu \, \sqrt{\frac{2}{\eps}} \; \dd \Btwo_t, \quad Y_0 = y \in \R,
\label{eqn:23} 
\end{equation}
in which $\Btwo$ is a standard Brownian motion on $(\Omega, {\cal F}, \P, {\bf F})$.  Suppose $\Bone$ and $\Btwo$ are correlated with (constant) coefficient $\rho_{12} \in (-1, 1)$.

Under its invariant distribution ${\cal N}(m, \nu^2)$, the autocorrelation of $Y$ is given by
\begin{equation}
\E \left[ (Y_t - m) (Y_s - m) \right] = \nu^2 \, e^{- {|t - s|}/{\eps}}.
\label{eqn:24} 
\end{equation}
Therefore, the process decorrelates exponentially fast on the time scale $\eps$; thus, we refer to $Y$ as the fast volatility factor.

The second factor $Z$ driving the volatility of the risky asset's price process is a slowly varying diffusion process.  We obtain this diffusion by applying the time change $t \to \del \cdot t$ to a given diffusion process:
\begin{equation}
\dd \widetilde Z_t = g( \widetilde Z_t) \, dt + h( \widetilde Z_t) \, \dd \tilde B_t,
\label{eqn:25} 
\end{equation}
in which $0 < \del \ll 1$ and $\tilde B$ is a standard Brownian motion.  The coefficients $g$ and $h$ are smooth and at most linearly growing at infinity, so (\ref{eqn:25}) has a unique strong solution.
Under the time change $t \to \del \cdot t$, define $Z_t = \widetilde Z_{\del \cdot t}$.  Then, the dynamics of $Z$ are given by
\begin{equation}
\dd Z_t = \del \, g(Z_t) \,dt +  h(Z_t) \, \dd \tilde B_{\del \cdot t}, \quad Z_0 = z \in \R.
\label{eqn:26} 
\end{equation}
In distribution, we can write these dynamics as
\begin{equation}
\dd Z_t = \del \, g(Z_t) \,dt + \sqrt{\del} \, h(Z_t) \, \dd\Bthr_t, \quad Z_0 = z \in \R,
\label{eqn:27} 
\end{equation}
in which $\Bthr$ is a standard Brownian motion on $(\Omega, {\cal F}, \P, {\bf F})$.  Suppose $\Bone$ and $\Bthr$ are correlated with (constant) coefficient $\rho_{13} \in [-1, 1]$.  Similarly, suppose $\Btwo$ and $\Bthr$ are correlated with (constant) coefficient $\rho_{23} \in [-1, 1]$. To ensure that the covariance matrix of the Brownian motions is positive semi-definite, we impose the following condition on the $\rho$'s:
\begin{equation}
1+2 \rho_{12} \rho_{13} \rho_{23} - \rho_{12}^2 -\rho_{13}^2 - \rho_{23}^2 \geq 0.
\label{eqn:rhos}
\end{equation}

Let $W_t$ be the wealth at time $t$ of the individual, and let $\pi_t$ be the amount that the decision maker invests in the risky asset at that time.  It follows that the amount invested in the riskless asset is $W_t - \pi_t$.  We assume that the individual consumes at a constant rate $c > 0$.  Therefore, the wealth process follows
\begin{equation}
\dd W_t = [rW_t + (\mu - r) \pi_t  - c] \, dt + f(Y_t, Z_t) \, \pi_t \, \dd \Bone_t, \label{eqn:28} 
\end{equation}
and we suppose that initial wealth is non-negative; that is, $W_0 = w \ge 0$.

By {\it lifetime ruin}, we mean that the individual's wealth reaches zero before she dies.  Define the corresponding hitting time by $\tau_0 := \inf\{t \ge 0: W_t \le 0 \}$.  Let $\tau_d$ denote the random time of death of the individual, which is independent of the Brownian motions.  We assume that $\tau_d$ is exponentially distributed with parameter $\lambda$ (that is, with expected time of death equal to $1/\lambda$); this parameter is also known as the {\it hazard rate}, or, {\it force of mortality}.

\citet{MR2328666} minimize the probability of lifetime ruin with varying hazard rate and show that by updating the hazard rate each year and treating it as a constant, the individual can quite closely obtain the minimal probability of ruin when the true hazard rate is Gompertz.  Specifically, at the beginning of each year, set $\la$ equal to the inverse of the individual's life expectancy at that time.  Compute the corresponding optimal investment strategy as given below, and apply that strategy for the year.  According to the work of \citet{MR2328666}, this scheme results in a  probability of ruin close to the minimum probability of ruin.   Therefore, there is no significant loss of generality to assume that the hazard rate is constant and revise its estimate each year.

Denote the minimum probability of lifetime ruin by $\psi(w, y, z)$, in which the arguments $w$, $y$, and $z$ indicate that one conditions on the individual possessing wealth $w$ at the current time with the two factors $Y$ and $Z$ taking the values $y$ and $z$, respectively, then.  Thus, $\psi$ is the minimum probability that $\tau_0 < \tau_d$, in which one minimizes with respect to admissible investment strategies $\pi$.  A strategy $\pi$ is {\it admissible} if it is ${\cal F}_t$-progressively measurable, and if it satisfies the integrability condition $\int_0^t \pi_s^2 \, ds < \infty$ almost surely for all $t \ge 0$.  Thus, $\psi$ is formally defined by
\begin{equation}
\psi(w, y, z) = \inf_{\pi} \P^{w, y, z} \left[\tau_0 < \tau_d \right].
\label{eqn:29} 
\end{equation}
Here, $\P^{w, y, z}$ indicates the probability conditional on $W_0 = w$, $Y_0 = y$, and $Z_0 = z$.  
Note that if $w \ge c/r$, then $\psi(w, y, z) = 0$ because the individual can invest $c/r$ of her wealth in the riskless asset and generate a rate of income equal to $c$, which exactly covers her consumption.  Therefore, we effectively only need to determine the minimum probability of lifetime ruin and corresponding optimal investment strategy on the domain ${\bf D} := \{ (w, y, z) \in \R^3: w \in [0, c/r] \}$.

\section{ Computing the Minimum Probability of Lifetime Ruin}

\subsection{ A Related Optimal Controller-Stopper Problem}

In this section, we present an optimal controller-stopper problem whose solution $\hat{\psi}$ is the Legendre dual of the minimum  probability of ruin $\psi$.  It is not clear {\it a priori} that the value function $\psi$ is convex or smooth due to the implicit dependent on the initial values of the state variable. By passing to the controller-stopper problem, however, we can obtain the regularity of $\hat{\psi}$ in more simply, which, in turn, provides an intermediate tool in the proof of regularity of $\psi$. The dual relationship and the analysis of the controller-stopper problem are, therefore, crucial and worth investigating.

  First, note that we can represent the three Brownian motions from Section 2 as follows:  given $\Bone$, $\Btwo$, and $\Bthr$, define $\widetilde{B}^{(1)}$, $\widetilde{B}^{(2)}$, and $\widetilde{B}^{(3)}$ via the following invertible system of equations:
\begin{equation}
\begin{split}
\Bone_t &= \widetilde{B}^{(1)}_t, \\
\Btwo_t &= \rho_{12} \, \widetilde{B}^{(1)}_t + \sqrt{1 - \rho^2_{12}} \, \widetilde{B}^{(2)}_t, \\
\Bthr_t &= \rho_{13} \, \widetilde{B}^{(1)}_t + \frac{\rho_{23}- \rho_{12} \rho_{13} }{\sqrt{1 - \rho^2_{12}}} \, \widetilde{B}^{(2)}_t + \frac{\sqrt{(1 - \rho^2_{12})(1 - \rho^2_{13}) - (\rho_{23} - \rho_{12} \rho_{13})^2} }{ \sqrt{1 - \rho^2_{12}}} \, \widetilde{B}^{(3)}_t.
\end{split}
\label{eqn:41} 
\end{equation}
One can show that $\widetilde{B}^{(1)}$, $\widetilde{B}^{(2)}$, and $\widetilde{B}^{(3)}$ thus defined are three {\it independent} standard Brownian motions on $(\Omega, {\cal F}, \P, {\bf F})$. Also notice that condition (\ref{eqn:rhos}) on the $\rho$'s guarantees that the expression under the square root in the coefficient of $\widetilde{B}^{(3)}_t$ is non-negative. 

Next, define the controlled process $X^\gam$ by
\begin{equation}
\dd X^\gam_t = -(r - \la) \, X^\gam_t \, dt - \frac{\mu - r }{ f(Y_t, Z_t)} \, X^\gam_t \, \dd \widetilde{B}^{(1)}_t + \gam^{(2)}_t \dd \widetilde{B}^{(2)}_t + \gam^{(3)}_t \dd \widetilde{B}^{(3)}_t,  \quad X_0 = x > 0,
\label{eqn:42} 
\end{equation}
in which $\gam = \left(\gam^{(2)}, \gam^{(3)} \right)$ is the control, and $Y$ and $Z$ are given in (\ref{eqn:23}) and (\ref{eqn:27}), respectively.

For $x > 0$, define the function $\hat \psi$ by
\begin{equation}
\hat \psi(x, y, z) = \inf_\tau \sup_\gam \E^{x, y, z} \left[ \int_0^\tau e^{-\la t} c \, X^\gam_t \, dt  + e^{-\la \tau} \min\left( (c/r)X^\gam_\tau, 1 \right) \right].
\label{eqn:43} 
\end{equation}
$\hat \psi$ is the value function for an optimal controller-stopper problem.  Indeed, the controller chooses among processes $\gam$ in order to maximize the discounted running ``penalty'' to the stopper given by $c \, X^\gam_t$ in (\ref{eqn:43}).  On the other hand, the stopper chooses the time to stop the game in order to minimize the penalty but has to incur the terminal cost of $\min\left( (c/r)X^\gam_\tau, 1 \right)$, discounted by $e^{-\la \tau}$ when she stops.

\citet{Bayraktar_Young-fs09} consider a controller-stopper problem that is mathematically similar to the one in this paper; see that paper for details of the following assertions--specifically, see Theorem 2.4 and its proof.  One can show that the controller-stopper problem has a continuation region given by $\{ (x, y, z): 0 \le x_{c/r}(y, z) \le x \le x_0(y, z) \}$ for some functions $0 \le x_{c/r}(y, z) \le r/c \le x_0(y, z)$ with $(y, z) \in \R^2$.  Thus, if $x \le x_{c/r}(y, z)$, we have $\hat \psi(x, y, z) = (c/r) \, x$, and if $x \ge x_0(y, z)$, we have $\hat \psi(x, y, z) = 1$.
Moreover, $\hat \psi$ is non-decreasing and concave with respect to $x$ on $\R^+$ (increasing and strictly concave in the continuation region) and is the unique classical solution of the following free-boundary problem on $\left[ x_{c/r}(y, z), \, x_0(y, z) \right]$:
\begin{equation}
\begin{cases} & cx + \left( \frac{1}{ \eps} \, \L_0 + \frac{1 }{ \sqrt{\eps}} \, \L_1 + \L_2  + \sqrt{\del} \, \M_1 + \del \, \M_2 + \sqrt{\frac{\del }{ \eps}} \, \M_3  \right) \hat \psi + NL^{\eps, \del}  = 0; \\
&\hat \psi(x_{c/r}(y, z), y, z) = \frac{c }{ r} \; x_{c/r}(y, z), \quad \hat \psi_x(x_{c/r}(y, z), y, z) = \frac{c }{ r}; \\
&\hat \psi(x_0(y, z), y, z) = 1, \quad \hat \psi_x(x_0(y, z), y, z) = 0;
\end{cases} 
\label{eqn:44} 
\end{equation}
in which
\begin{eqnarray}
\L_0 v &=& (m- y) \, v_y + \nu^2 \, v_{yy}, \label{eqn:45}\\
\L_1 v &=& - \rho_{12} \, \frac{\mu - r }{ f(y, z)} \, \nu \, \sqrt{2} \, x \, v_{xy},\label{eqn:46}\\
\L_2 v &=& - \la \, v - (r - \la) \, x \, v_x + \frac{1}{2} \left( \frac{\mu - r }{ f(y, z)} \right)^2 \, x^2 \, v_{xx},\label{eqn:47}\\
\M_1 v &=& - \rho_{13} \, \frac{\mu - r }{ f(y, z)} \, h(z) \, x \, v_{xz},\label{eqn:48}\\
\M_2 v &=& g(z) \, v_z + \frac{1 }{ 2} \, h^2(z) \, v_{zz},\label{eqn:49}\\
\M_3 v &=& \rho_{23} \, \nu \, \sqrt{2} \, h(z) \, v_{yz},
\label{eqn:410} 
\end{eqnarray}
and
\begin{equation}
\begin{split}
NL^{\eps, \del} &= \sup_{\gam} \left[\rule{0cm}{0.9cm} \frac{1 }{ 2} \left( \left( \gam^{(2)} \right)^2 + \left( \gam^{(3)} \right)^2 \right) \hat \psi_{xx}  \right. \\
& \; \qquad \quad + \, \gam^{(2)} \left(  \nu \, \sqrt{\frac{2 }{ \eps}}  \, \sqrt{1 - \rho^2_{12}} \; \hat \psi_{xy} + \sqrt{\del} \, h(z) \, {\frac{\rho_{23} - \rho_{12} \rho_{23} }{ \sqrt{1 - \rho^2_{12}}}} \; \hat \psi_{xz} \right) \\
& \; \qquad \quad \left. + \, \gam^{(3)} \, \sqrt{\del} \, h(z) \, \frac{\sqrt{(1 - \rho^2_{12})(1 - \rho^2_{13}) - (\rho_{23} - \rho_{12} \rho_{13})^2} }{ \sqrt{1 - \rho^2_{12}}} \; \hat \psi_{xz} 
\right]\rule{0cm}{0.9cm}.
\end{split}
\label{eqn:411} 
\end{equation}
Because $\hat \psi$ is concave with respect to $x$, we can express $NL^{\eps, \del}$ as follows:
\begin{equation}
NL^{\eps, \del} = - \frac{1}{\eps} \, \nu^2 \left(1 - \rho^2_{12} \right) \frac{\hat \psi^2_{xy} }{ \hat \psi_{xx}} - \frac{1 }{ 2} \, \del \, h^2(z) \left( 1- \rho^2_{13} \right) \frac{\hat \psi^2_{xz} }{ \hat \psi_{xx}} - \nu \sqrt{2} \, \sqrt{\frac{\del }{ \eps}} \; h(z) \left( \rho_{23} - \rho_{12} \rho_{13} \right) \frac{\hat \psi_{xy} \hat \psi_{xz} }{ \hat \psi_{xx}}. 
\label{eqn:412} 
\end{equation}

\subsection{  Convex Legendre Dual of $\hp$}

Since $\hat \psi$ is strictly concave with respect to $x$ in its continuation region (which corresponds to wealth lying in $[0, c/r]$), we can define its convex dual $\Psi$ by the Legendre transform:  for $(w, y, z) \in {\bf D} = \{ (w, y, z) \in \R^3: w \in [0, c/r] \}$,
\begin{equation}
\Psi(w, y, z) = \max_x \left( \hat \psi(x, y, z) - wx \right). \label{eqn:413} 
\end{equation}
In this section, we show that the convex dual $\Psi$ is the minimum probability of lifetime ruin; then, in the next section, we asymptotically expand $\hat \psi$ in powers of $\sqrt{\eps}$ and $\sqrt{\del}$.

{\theorem 
\label{thm:41}
$\Psi$ equals the minimum probability of lifetime ruin $\psi$ on $\bf D$, and the investment policy $\pv$ given in feedback form by $\pv_t = \pv(W^*_t, Y_t, Z_t)$ is an optimal policy, in which $W^*$ is the optimally controlled wealth $($that is, wealth controlled by $\pv)$ and the function $\pv$ is given by
\begin{equation}
\pv(w, y, z) = - \frac{\mu - r }{ f^2(y, z)} \, \frac{\psi_w }{ \psi_{ww}} - \rho_{12} \, \sqrt{\frac{2 }{ \eps}} \, \frac{\nu }{ f(y, z)} \, \frac{\psi_{wy} }{ \psi_{ww}} - \rho_{13} \, \sqrt{\del} \; \frac{ h(z) }{f(y, z)} \, \frac{\psi_{wz} }{ \psi_{ww}} \, ,
\label{eqn:414} 
\end{equation}
in which the right-hand side of $(\ref{eqn:414})$ is evaluated at $(w, y, z)$.}

\pf  From (\ref{eqn:413}), it follows that the critical value $x^*$ solves $w = \hat \psi_x(x, y, z)$; thus, given $w$, we have $x^* = I(w, y, z)$, in which $I$ is the inverse function of $\hat \psi_x$ with respect to $x$.  Therefore, $\Psi(w, y, z) = \hat \psi(I(w, y, z), y, z) - w I(w, y, z)$.  By differentiating this expression of $\Psi$ with respect to $w$, we obtain $\Psi_w(w, y, z) = \hat \psi_x(I(w, y, z), y, z) I_w(w, y, z) - I(w, y, z) - w I_w(w, y, z) = - I(w, y, z)$; thus, $x^* = - \Psi_w(w, y, z)$. Similarly, we obtain (with $w = \hat \psi_x(x, y, z))$ the following expressions:
\begin{equation}
\hat \psi_{xx}(x, y, z) = - \, \frac{1 }{ \Psi_{ww}(w, y, z)},
\label{eqn:415} 
\end{equation}
\begin{equation}
\hat \psi_y(x, y, z) = \Psi_y(w, y, z),
\label{eqn:416} 
\end{equation}
\begin{equation}
\hat \psi_z(x, y, z) = \Psi_z(w, y, z),
\label{eqn:417} 
\end{equation}
\begin{equation}
\hat \psi_{yy}(x, y, z) = \Psi_{wy}(w, y, z) \, \hat \psi_{xy}(x, y, z) + \Psi_{yy}(w, y, z),
\label{eqn:418} 
\end{equation}
\begin{equation}
\hat \psi_{zz}(x, y, z) = \Psi_{wz}(w, y, z) \, \hat \psi_{xz}(x, y, z) + \Psi_{zz}(w, y, z),
\label{eqn:419} 
\end{equation}
\begin{equation}
\hat \psi_{xy}(x, y, z) = \Psi_{wy}(w, y, z) \, \hat \psi_{xx}(x, y, z),
\label{eqn:420} 
\end{equation}
\begin{equation}
\hat \psi_{xz}(x, y, z) = \Psi_{wz}(w, y, z) \, \hat \psi_{xx}(x, y, z),
\label{eqn:421} 
\end{equation}
and
\begin{equation}
\hat \psi_{yz}(x, y, z) = \Psi_{wy}(w, y, z) \, \Psi_{wz}(w, y, z) \, \hat \psi_{xx}(x, y, z) + \Psi_{yz}(w, y, z).
\label{eqn:422} 
\end{equation}
By substituting $x^* = - \Psi_w(w, y, z)$ into the free-boundary problem for $\hat \psi$, namely (\ref{eqn:44}), one can show  that $\Psi$ uniquely solves the following boundary-value problem on $\bf D$:
\begin{equation}
\begin{cases}
&\min_\beta \D^\beta v(w, y, z) = 0; \\
&v(0, y, z) = 1, \quad v(c/r, y, z) = 0.
\end{cases}
\label{eqn:423} 
\end{equation}
where the differential operator $\D^\beta$ is given by
\begin{equation}
\begin{split}
\D^\beta v &= - \la \, v + (rw + (\mu - r) \beta - c) \, v_w + \frac{1}{ \eps} \, (m-y) \, v_y + \del \, g(z) \, v_z \\
& \quad + \frac{1}{2} f^2(y,z) \, \beta^2 \, v_{ww} + \frac{1}{ \eps} \, \nu^2 \, v_{yy} + \frac{1}{2} \, \del \, h^2(z) \, v_{zz} + \rho_{12} \, f(y, z) \, \beta \, \nu \, \sqrt{\frac{2 }{ \eps}} \, v_{wy} \\
& \quad + \rho_{13} \, f(y, z) \, \beta \, \sqrt{\del} \, h(z) \, v_{wz} + \rho_{23} \, \sqrt{2} \, \nu \, \sqrt{\frac{\del}{ \eps}} \, h(z) \, v_{yz}.
\end{split}
\label{eqn:31} 
\end{equation}

Observe that $\Psi$ is strictly convex in $w$ because $\hat{\psi}$ is strictly concave in $x$ in its continuation region which corresponds to $\bf D$ in the original space. Since $\Psi$ is strictly convex with respect to $w$, the optimal policy $\pv$ in (\ref{eqn:423}) is given by the first-order necessary condition, which results in the expression in (\ref{eqn:414}).  Now, using a standard verification theorem we deduce that $\Psi$ is the minimum probability of lifetime ruin $\psi$.  \qed

Theorem \ref{thm:41} demonstrates the strong connection between $\hp$ and $\psi$, namely that they are dual via the Legendre transform.  (As an aside, if we have $\psi$, we can obtain $\hp$ via $\hp(x, y, z) = \min_w \left( \psi(w, y, z) + wx \right)$.)  Therefore, if we have $\hp$, then we obtain the minimum probability of ruin $\psi$ via (\ref{eqn:413}).  More importantly, we get the optimal investment strategy $\pv$ via (\ref{eqn:414}).  As a corollary to Theorem \ref{thm:41}, we have the following expression for $\pv$ in terms of the dual variable $x$.

{\corollary
\label{cor:42}
In terms of the dual variable $x$, the optimal investment strategy $\pv$ is given by $\pv_t = \hat{\pi}^*(X^*_t, Y_t, Z_t)$, in which $X^*$ is the optimally controlled process $X$, and
\begin{equation}
\hat{\pi}^*(x, y, z) = - \frac{\mu - r }{ f^2(y, z)} \, x \, \hp_{xx} + \rho_{12} \, \sqrt{\frac{2 }{ \eps}} \, \frac{\nu }{ f(y, z)} \, \hp_{xy} + \rho_{13} \, \sqrt{\del} \; \frac{h(z) }{ f(y, z)} \, \hp_{xz},
\label{eqn:424} 
\end{equation}
with the right-hand side of $(\ref{eqn:424})$ evaluated at $(x, y, z)$.}

\pf Let $w = \hp_x(x, y, z)$ in (\ref{eqn:414}) and simplify the right-hand side via equations (\ref{eqn:415})-(\ref{eqn:422}) to obtain (\ref{eqn:424}).  \qed


\section{Asymptotic Approximation of the Minimum Probability of Lifetime Ruin}

In this section, we asymptotically expand $\hp$, the Legendre transform of the minimum probability of ruin, in powers of $\sqrt{\eps}$ and $\sqrt{\del}$. (A parallel analysis of expanding the Legendre transform of the value function of the utility maximization problem was carried out in \citet{JSMFIN03}.)  We expand $\hat \psi$ instead of $\psi$ because if one were to do the latter, then one would note that each term in the expansion solves a {\it non-linear} differential equation.  The differential equation for the zeroth-order term has a closed-form solution; however, none of the differential equations for the higher-order terms does.  What this fact implies is that to solve any of these non-linear differential equations, one would have to assume that it has a convex solution, determine the corresponding linear free-boundary problem for the concave dual, solve this free-boundary problem, then invert the solution numerically, as in equation \eqref{eqn:413}.  Note that one would have to perform this procedure for {\it each} higher-order term in the expansion.

By contrast, when we expand $\hat \psi$, each term solves a {\it linear} differential equation, as we show below.  We explicitly solve these linear differential equations, then invert the approximation using \eqref{eqn:413} {\it once} to obtain an approximation for the minimum probability of lifetime ruin $\psi$.  Note that the resulting approximation of $\psi$ is not guaranteed to be a probability, that is, to lie in the interval $[0, 1]$; however, our numerical experiments show that this is not a problem for the values of the parameters we consider.  See \cite{Fprob} for an example of approximating a probability that solves a linear differential equation.

To begin, expand $\hp$ and the free boundaries in powers of $\sqrt{\del}$:
\begin{equation}
\hp = \hpz + \sqrt{\del} \, \hpo + \del \, \hpt + \cdots,
\label{eqn:425} 
\end{equation}
\begin{equation}
x_{c/r}(y, z) = x_{c/r, 0}(y, z) + \sqrt{\del} \, x_{c/r, 1}(y, z) + \del \, x_{c/r, 2}(y, z) + \cdots,
\label{eqn:426} 
\end{equation}
and
\begin{equation}
x_{0}(y, z) = x_{0, 0}(y, z) + \sqrt{\del} \, x_{0, 1}(y, z) + \del \, x_{0, 2}(y, z) + \cdots.
\label{eqn:427} 
\end{equation}
Insert the expression in (\ref{eqn:425}) into $NL^{\eps, \del}$ in (4.12) to obtain the following expansion in powers of $\sqrt{\del}$:
\begin{equation}
\begin{split}
NL^{\eps, \del} &= - \, \frac{1 }{ \eps} \, \nu^2 \left( 1 - \rho^2_{12} \right) \, \frac{\hp^2_{0, xy} }{ \hp_{0, xx}} \\
& \quad + \sqrt{\del} \left[ \frac{1 }{ \eps} \, \nu^2 \left( 1 - \rho^2_{12} \right) \left( \left( \frac{\hp_{0, xy} }{ \hp_{0, xx}} \right)^2 \hp_{1, xx} - 2 \; \frac{\hp_{0, xy} }{ \hp_{0, xx}} \; \hp_{1, xy} \right) \right. \\
& \qquad \qquad  \left. - \, \sqrt{\frac{2 }{ \eps}} \; \nu \, h(z) (\rho_{23} - \rho_{12} \rho_{13}) \frac{\hp_{0, xy} \, \hp_{0, xz} }{ \hp_{0, xx}} \right] + {\cal O}(\del).
\end{split}
\label{eqn:428} 
\end{equation}

Keeping terms up to $\sqrt{\del}$, we expand the free-boundary conditions in (\ref{eqn:44}) as
\begin{equation}
\begin{split}
& \hpz(x_{c/r, 0}(y, z), y, z) + \sqrt{\del} \left[ x_{c/r, 1}(y, z) \, \hp_{0, x}(x_{c/r, 0}(y, z), y, z) + \hpo(x_{c/r, 0}(y, z), y, z) \right]\\
& \quad = \frac{c }{ r} \left(  x_{c/r, 0}(y, z) + \sqrt{\del} \; x_{c/r, 1}(y, z) \right),
\end{split}
\label{eqn:429} 
\end{equation}
\begin{equation}
\begin{split}
& \hp_{0, x}(x_{c/r, 0}(y, z), y, z) + \sqrt{\del} \left[ x_{c/r, 1}(y, z) \, \hp_{0, xx}(x_{c/r, 0}(y, z), y, z) + \hp_{1, x}(x_{c/r, 0}(y, z), y, z) \right]\\
& \quad = \frac{c }{ r},
\end{split}
\label{eqn:430} 
\end{equation}
\begin{equation}
\hpz(x_{0, 0}(y, z), y, z) + \sqrt{\del} \left[ x_{0, 1}(y, z) \, \hp_{0, x}(x_{0, 0}(y, z), y, z) + \hpo(x_{0, 0}(y, z), y, z) \right] = 1,
\label{eqn:431} 
\end{equation}
and
\begin{equation}
\hp_{0, x}(x_{0, 0}(y, z), y, z) + \sqrt{\del} \left[ x_{0, 1}(y, z) \, \hp_{0, xx}(x_{0, 0}(y, z), y, z) + \hp_{1, x}(x_{0, 0}(y, z), y, z) \right] = 0.
\label{eqn:432} 
\end{equation}
We begin by approximating $\hpz$ and the free boundaries $x_{c/r, 0}$ and $x_{0, 0}$.  Then, we use the boundaries $x_{c/r, 0}$ and $x_{0, 0}$ as {\it fixed} boundaries to determine $\hpo$.  As one can see from equations (\ref{eqn:429})-(\ref{eqn:432}), this fixing of the boundaries introduces an ${\cal O}(\sqrt{\del})$-error into $\hpo$ in ${\cal O}(\sqrt{\del})$-neighborhoods of $x_{c/r, 0}$ and $x_{0, 0}$.

\subsection*{Terms of order $\del^0$} By inserting (\ref{eqn:425})-(\ref{eqn:428}) into (\ref{eqn:44}) and collecting terms of order $\del^0$, we obtain the following free-boundary problem:
\begin{equation}
\begin{cases}&cx + \left( \frac{1 }{ \eps} \, \L_0 + \frac{1 }{ \sqrt{\eps}} \, \L_1 + \L_2  \right) \hpz - \, \frac{1 }{ \eps} \, \nu^2 \left( 1 - \rho^2_{12} \right) \, \frac{\hp^2_{0, xy} }{ \hp_{0, xx}}  = 0; \\
&\hpz(x_{c/r, 0}(y, z), y, z) = \frac{c }{ r} \, x_{c/r, 0}(y, z), \quad \hp_{0, x}(x_{c/r,0}(y, z), y, z) = \frac{c }{ r}; \\
&\hpz(x_{0, 0}(y, z), y, z) = 1, \quad \hp_{0, x}(x_{0, 0}(y, z), y, z) = 0.
\end{cases}
\label{eqn:433} 
\end{equation}

\medskip

\subsection*{Terms of order $\sqrt{\del}$}  Similarly, by comparing terms of order $\sqrt{\del}$ and using $x_{c/r, 0}$ and $x_{0, 0}$ as fixed boundaries for $\hpo$, we obtain the following boundary-value problem:
\begin{equation}
\begin{cases}
&\left( \frac{1 }{ \eps} \, \L_0 + \frac{1 }{ \sqrt{\eps}} \, \L_1 + \L_2  \right) \hpo + \left( \M_1 + \frac{1 }{ \sqrt{\eps}} \, \M_3 \right) \hpz \\
& \quad + \, \frac{1 }{ \eps} \, \nu^2 \left( 1 - \rho^2_{12} \right) \left( \left( \frac{\hp_{0, xy} }{ \hp_{0, xx}} \right)^2 \hp_{1, xx} - 2 \; \frac{\hp_{0, xy} }{ \hp_{0, xx}} \; \hp_{1, xy} \right) \\
& \quad - \sqrt{\frac{2}{ \eps}} \; \nu \, h(z) (\rho_{23} - \rho_{12} \rho_{13}) \frac{\hp_{0, xy} \, \hp_{0, xz} }{ \hp_{0, xx}} = 0; \\
&\hpo(x_{c/r, 0}(y, z), y, z) = 0, \quad \hpo(x_{0,0}(y, z), y, z) = 0.
\end{cases}
\label{eqn:434} 
\end{equation}

\medskip

Next, we expand the solutions of (\ref{eqn:433}) and (\ref{eqn:434}) in powers of $\sqrt{\eps}$:
\begin{equation}
\hpz(x, y, z) = \hp_{0, 0}(x, y, z) + \sqrt{\eps} \, \hp_{0, 1}(x, y, z) + \eps \, \hp_{0, 2}(x, y, z) + \cdots,
\label{eqn:435} 
\end{equation}
and
\begin{equation}
\hpo(x, y, z) = \hp_{1, 0}(x, y, z) + \sqrt{\eps} \, \hp_{1, 1}(x, y, z) + \eps \, \hp_{1, 2}(x, y, z) + \cdots.
\label{eqn:436} 
\end{equation}
Similarly, expand the free boundaries $x_{c/r, 0}$ and $x_{0, 0}$ in powers of $\sqrt{\eps}$:
\begin{equation}
x_{c/r, 0}(y, z) = x_{c/r, 0, 0}(y, z) + \sqrt{\eps} \, x_{c/r, 0, 1}(y, z) + \eps \, x_{c/r, 0, 2}(y, z) + \cdots,
\label{eqn:437} 
\end{equation}
and
\begin{equation}
x_{0, 0}(y, z) = x_{0, 0, 0}(y, z) + \sqrt{\eps} \, x_{0, 0, 1}(y, z) + \eps \, x_{0, 0, 2}(y, z) + \cdots.
\label{eqn:438} 
\end{equation}
Substitute (\ref{eqn:435}) and (\ref{eqn:436}) into (\ref{eqn:433}) and (\ref{eqn:434}), respectively, and collect terms of the same order of $\sqrt{\eps}$.  As discussed earlier, we determine the free boundaries $x_{c/r, 0, 0}(y, z)$ and $x_{0, 0, 0}(y, z)$ via a free-boundary problem for $\hp_{0, 0}$; then, we use these boundaries as the {\it fixed} boundaries for $\hp_{0, 1}$ and $\hp_{1, 0}$.

\subsection*{Terms of order $1/\eps$ in $(\ref{eqn:433})$}  By matching terms of order $1/\eps$ in (\ref{eqn:433}), we obtain the following:
\begin{equation}
\L_0 \, \hp_{0, 0} - \,  \, \nu^2 \left( 1 - \rho^2_{12} \right) \, \frac{\hp^2_{0, 0, xy}}{ \hp_{0, 0, xx}}  = 0,
\label{eqn:439} 
\end{equation}
or equivalently
\begin{equation}
(m - y) \, \hp_{0, 0, y} + \nu^2 \, \hp_{0, 0, yy}  - \,  \, \nu^2 \left( 1 - \rho^2_{12} \right) \, \frac{\hp^2_{0, 0, xy} }{ \hp_{0, 0, xx}}  = 0.
\label{eqn:440} 
\end{equation}
We, therefore, look for an $\hp_{0,0}$ independent of $y$; otherwise, $\hp_{0,0}$ will experience exponential growth as $y$ goes to $\pm \infty$ \citep{MRsircar,Fouque:asymp}.  We also seek free boundaries $x_{c/r, 0, 0}$ and $x_{0, 0, 0}$ independent of $y$.

\subsection*{Terms of order $1/\sqrt{\eps}$ in $(\ref{eqn:433})$}  By matching terms of order $1/\sqrt{\eps}$ in (\ref{eqn:433}) and using the fact that $\hp_{0, 0, y} \equiv 0$, we obtain the following:
\begin{equation}
\L_0 \, \hp_{0, 1}  = 0.
\label{eqn:441} 
\end{equation}
Therefore, we look for an $\hp_{0, 1}$ independent of $y$; otherwise, $\hp_{0, 1}$ will experience exponential growth as $y$ goes to $\pm \infty$.

\medskip

\subsection*{Terms of order $\eps^0$ in $(\ref{eqn:433})$}  By matching terms of order $\eps^0$ in (\ref{eqn:433}) and using the fact that  $\hp_{0, 0, y} = \hp_{0, 1, y} \equiv 0$, we obtain the following Poisson equation (in $y$) for $\hp_{0, 2}$:
\begin{equation}
\L_0 \, \hp_{0, 2} = -cx - \L_2 \, \hp_{0, 0}.
\label{eqn:442} 
\end{equation}
The solvability condition for this equation requires that $cx + \L_2 \, \hp_{0, 0}$ be centered with respect to the invariant distribution ${\cal N}(m, \nu^2)$ of the Ornstein-Uhlenbeck process $Y$.  Specifically,
\begin{equation}
\left \langle cx + \L_2 \, \hp_{0, 0} \right \rangle = cx + \left \langle \L_2 \right \rangle \hp_{0, 0} = 0,
\label{eqn:443} 
\end{equation}
in which $\langle \cdot \rangle$ denotes averaging with respect to the distribution ${\cal N}(m, \nu^2)$:
\begin{equation}
\langle v \rangle = \frac{1 }{ \sqrt{2 \pi \nu^2}}\, \int_{-\infty}^\infty v(y) \, e^{- \frac{(y- m)^2 }{2 \nu^2}} \, \dd y.
\label{eqn:444} 
\end{equation}
In (4.43), the averaged operator $\left \langle \L_2 \right \rangle$ is defined by
\begin{equation}
\left \langle \L_2 \right \rangle v = -\la \, v - (r - \la) \, x \, v_x + \frac{1 }{ 2} \left( \frac{\mu - r }{ \sigma_*(z)} \right)^2 \, x^2 \, v_{xx},
\label{eqn:445} 
\end{equation}
in which $\sigma_*(z)$ is given by
\begin{equation}
\frac{1 }{ \sigma^2_*(z)} = \left \langle \frac{1 }{ f^2(y, z)} \right \rangle.
\label{eqn:446} 
\end{equation}

Thus, we have the following free-boundary problem for $\hp_{0, 0}$:
\begin{equation}
\begin{cases}
& cx -\la \, \hp_{0, 0} - (r - \la) \, x \, \hp_{0, 0, x} + s(z) \, x^2 \, \hp_{0, 0, xx} = 0 \, ; \\
& \hp_{0, 0}(x_{c/r, 0, 0}(z), z) = \frac{c }{ r} \, x_{c/r, 0, 0}(z), \quad \hp_{0, 0, x}(x_{c/r, 0, 0}(z), z) =\frac{c }{ r} \, ; \\
&\hp_{0, 0}(x_{0, 0, 0}(z), z) = 1, \quad \hp_{0, 0, x}(x_{0, 0, 0}(z), z) = 0.
\end{cases}
\label{eqn:447} 
\end{equation}
with $s(z) = \frac{1 }{ 2} \left( \frac{\mu - r }{ \sigma_*(z)} \right)^2$.  The general solution of the differential equation in (\ref{eqn:447}) is given by
\begin{equation}
\hp_{0, 0}(x, z) = D_1(z) \, x^{B_1(z)} + D_2(z) \, x^{B_2(z)} + \frac{c }{ r} \, x,
\label{eqn:448} 
\end{equation}
in which
\begin{equation}
B_1(z) = \frac{1 }{2 s(z)} \left[ \left(r - \lambda + s(z) \right) + \sqrt{ \left(r - \lambda + s(z) \right)^2 + 4 \lambda s(z)} \right] > 1,
\label{eqn:449} 
\end{equation}
and
\begin{equation}
B_2(z) = \frac{1 }{ 2 s(z)} \left[ \left(r - \lambda + s(z) \right) - \sqrt{ \left(r - \lambda + s(z) \right)^2 + 4 \lambda s(z)} \right] < 0.
\label{eqn:450} 
\end{equation}
We determine $D_1$ and $D_2$ from the free-boundary conditions.

The free-boundary conditions imply that
\begin{equation}
D_1(z) \, x_{c/r, 0, 0}(z)^{B_1(z)} + D_2(z) \, x_{c/r, 0, 0}(z)^{B_2(z)} + \frac{c }{ r} \, x_{c/r, 0, 0}(z) = \frac{c }{ r} \, x_{c/r, 0, 0}(z),
\label{eqn:451} 
\end{equation}
\begin{equation}
D_1(z) \, B_1(z) \, x_{c/r, 0, 0}(z)^{B_1(z) - 1} + D_2(z) \, B_2(z) \, x_{c/r, 0, 0}(z)^{B_2(z) - 1} + \frac{c}{r} = \frac{c}{r},
  \label{eqn:452} 
\end{equation}
\begin{equation}
D_1(z) \, x_{0, 0, 0}(z)^{B_1(z)} + D_2(z) \, x_{0, 0, 0}(z)^{B_2(z)} + \frac{c}{r} \, x_{0, 0, 0}(z) = 1,
\label{eqn:453} 
\end{equation}
and
\begin{equation}
D_1(z) \, B_1(z) \, x_{0, 0, 0}(z)^{B_1(z) - 1} + D_2(z) \, B_2(z) \, x_{0, 0, 0}(z)^{B_2(z) - 1} + \frac{c}{r} = 0,
\label{eqn:454} 
\end{equation}
which gives us four equations to determine the four unknowns $D_1$, $D_2$, $x_{c/r, 0, 0}$, and $x_{0, 0, 0}$.  Indeed, the solution to these equations is
\begin{equation}
D_1(z) = - \, \frac{1 }{ B_1(z) - 1} \left( \frac{c}{r} \cdot \frac{B_1(z) - 1}{ B_1(z)}  \right)^{B_1(z)},
\label{eqn:455} 
\end{equation}
\begin{equation}
D_2(z) \equiv 0,
\label{eqn:456} 
\end{equation}
\begin{equation}
x_{c/r, 0, 0}(z) \equiv 0,
\label{eqn:457} 
\end{equation}
and
\begin{equation}
x_{0, 0, 0}(z) =  \frac{B_1(z)}{ B_1(z) - 1} \cdot \frac{r }{ c}.
\label{eqn:458} 
\end{equation}
It follows that
\begin{equation}
\hp_{0, 0}(x, z) =  - \, \frac{1}{ B_1(z) - 1} \left( \frac{c}{r} \cdot \frac{B_1(z) - 1 }{ B_1(z)} \cdot x  \right)^{B_1(z)} + \frac{c}{r} \, x.
\label{eqn:459} 
\end{equation}

\subsection*{Terms of order $\sqrt{\eps}$ in $(\ref{eqn:433})$}  By matching terms of order $\sqrt{\eps}$ in (\ref{eqn:433}) and using the fact that  $\hp_{0, 0, y} = \hp_{0, 1, y} = 0$, we obtain the following Poisson equation (in $y$) for $\hp_{0, 3}$:
\begin{equation}
\L_0 \, \hp_{0, 3} = - \L_1 \, \hp_{0, 2} - \L_2 \, \hp_{0, 1}.
\label{eqn:460} 
\end{equation}
As above, the solvability condition for this equation requires that
\begin{equation}
\left \langle \L_1 \, \hp_{0, 2} + \L_2 \, \hp_{0, 1} \right \rangle = 0,
\label{eqn:461}
\end{equation}
in which
\begin{equation}
\hp_{0, 2}(x, z) = \L_0^{-1} \left( - cx - \L_2 \, \hp_{0, 0} \right).
\label{eqn:462} 
\end{equation}
It follows that $\hp_{0, 1}$ solves
\begin{equation}
\left \langle \L_2 \right \rangle \hp_{0, 1} = \left \langle \L_1 \L_0^{-1} \left( cx + \L_2 \, \hp_{0, 0} \right) \right \rangle.
\label{eqn:463} 
\end{equation}
Recall that we impose the (fixed) boundary conditions $\hp_{0,1}(x_{c/r, 0, 0}(z), z) = 0$ and \hfill \break $\hp_{0,1}(x_{0, 0, 0}(z), z) = 0$ at $x_{c/r, 0, 0}(z) \equiv 0$ and $x_{0, 0, 0}(z) =  \frac{B_1(z) }{ B_1(z) - 1} \cdot \frac{r }{ c}$.

From (\ref{eqn:462}), it is straightforward to show that $\hp_{0, 2}$ can be expressed as follows:
\begin{equation}
\hp_{0, 2}(x, y, z) = - D_1(z) \, B_1(z) \, (B_1(z) - 1) \, x^{B_1(z)} \, \eta(y, z),
\label{eqn:464} 
\end{equation}
in which $\eta$ solves
\begin{equation}
(m-y) \eta_y + \nu^2 \, \eta_{yy} = \frac{1}{2}\left( \frac{\mu - r }{ f(y, z)} \right)^2 - \frac{1}{2}\left( \frac{\mu - r }{ \sigma^*(z)} \right)^2 = \frac{1}{2}\left( \frac{\mu - r }{ f(y, z)} \right)^2 - s(z).
\label{eqn:465} 
\end{equation}
It follows that the right-hand side of (\ref{eqn:463}) equals
\begin{equation}
\begin{split}
& - \rho_{12} \, (\mu - r) \, \nu \, \sqrt{2} \, D_1(z) \, B_1^2(z) \, (B_1(z) - 1) \, x^{B_1(z)} \, \left \langle \frac{\eta_y(y, z) }{ f(y, z)} \right \rangle \\
& = \rho_{12} \, (\mu - r) \, \sqrt{2} {\nu} \, D_1(z) \, B_1^2(z) \, (B_1(z) - 1) \, x^{B_1(z)} \, \left \langle \tilde F(y, z) \left(   \frac{1}{2}\left( \frac{\mu - r }{ f(y, z)} \right)^2 - s(z) \right) \right \rangle ,
\end{split}
\label{eqn:466} 
\end{equation}
in which $\tilde F$ is an antiderivative of $1/f$ with respect to $y$; that is,
\begin{equation}
\tilde F_y(y, z) = \frac{1 }{ f(y, z)}.
\label{eqn:467} 
\end{equation}
From (\ref{eqn:463}) and (\ref{eqn:466}), we obtain that $\hp_{0, 1}$ equals
\begin{equation}
\hp_{0, 1}(x, z) = \tilde D_1(z) \, x^{B_1(z)} + \tilde D_2(z) \, x^{B_2(z)} + A(z) \, x^{B_1(z)} \, \ln x,
\label{eqn:468} 
\end{equation}
in which $B_1$ and $B_2$ are given in (\ref{eqn:449}) and (\ref{eqn:450}), respectively,  and $A$ is given by
\begin{equation}
A(z) = \frac{\rho_{12} \, (\mu - r) \, \sqrt{2} \nu\, D_1(z) \, B_1^2(z) \, (B_1(z) - 1) }{ (2 B_1(z) - 1) \, s(z) - (r - \la)} \left \langle \tilde F(y, z) \left(   \frac{1}{2}\left( \frac{\mu - r }{ f(y, z)} \right)^2 - s(z) \right) \right \rangle. 
\label{eqn:469} 
\end{equation}
The functions $\tilde D_1$ and $\tilde D_2$ are given by the (fixed) boundary conditions at $x_{c/r, 0, 0}(z) \equiv 0$ and $x_{0, 0, 0}(z) =  \frac{B_1(z) }{ B_1(z) - 1} \cdot \frac{r}{c}$, from which it follows that
\begin{equation}
\begin{split}
\hp_{0, 1}(x, z) &= A(z) \, x^{B_1(z)} \, \left( \ln x - \ln \left(  \frac{B_1(z) }{ B_1(z) - 1} \cdot \frac{r}{c} \right) \right) \\
&= A(z) \, x^{B_1(z)} \,  \ln \left( x \cdot  \frac{B_1(z) - 1 }{ B_1(z)} \cdot \frac{c}{r} \right).
\end{split}
\label{eqn:470} 
\end{equation}

Next, we focus on (\ref{eqn:434}) to find $\hp_{1, 0}$, after which we will approximate $\hp$ by $\hp_{0, 0} + \sqrt{\eps} \, \hp_{0, 1} + \sqrt{\del} \, \hp_{1, 0}$.

\subsection*{Terms of order $1/\eps$ in $(\ref{eqn:434})$}  By matching terms of order $1/\eps$ in (\ref{eqn:434}), we obtain the following:
\begin{equation}
\L_0 \, \hp_{1, 0} = 0,
\label{eqn:471} 
\end{equation} 
from which it follows that $\hp_{1,0}$ is independent of $y$; otherwise, $\hp_{1,0}$ will experience exponential growth as $y$ goes to $\pm \infty$  \citep{Fouque:asymp}.

\subsection*{Terms of order $1/\sqrt{\eps}$ in $(\ref{eqn:434})$}  By matching terms of order $1/\sqrt{\eps}$ in (\ref{eqn:434}) and using the fact that $\hp_{1, 0, y} \equiv 0$, we obtain the following:
\begin{equation}
\L_0 \, \hp_{1, 1}  = 0.
\label{eqn:472} 
\end{equation}
Therefore, we look for an $\hp_{1, 1}$ independent of $y$; otherwise, $\hp_{1, 1}$ will experience exponential growth as $y$ goes to $\pm \infty$.

\subsection*{Terms of order $\eps^0$ in $(\ref{eqn:434})$}  By matching terms of order $\eps^0$ in (\ref{eqn:434}) and using the fact that  $\hp_{1, 0, y} = \hp_{1, 1, y} \equiv 0$, we obtain the following Poisson equation (in $y$) for $\hp_{1, 2}$:
\begin{equation}
\L_0 \, \hp_{1, 2} = - \L_2 \, \hp_{1, 0} +  \rho_{13} \,  \frac{\mu - r }{ f(y, z)} \, h(z) \, x \, \hp_{0, 0, xz}.
\label{eqn:473} 
\end{equation}
The solvability condition for this equation requires that
\begin{equation}
\left \langle - \L_2 \, \hp_{1, 0} +  \rho_{13} \,  \frac{\mu - r }{ f(y, z)} \, h(z) \, x \, \hp_{0, 0, xz} \right \rangle = 0,
\label{eqn:474} 
\end{equation}
or equivalently,
\begin{equation}
\left \langle \L_2 \right \rangle \hp_{1, 0} =  \rho_{13} \, \left \langle   \frac{\mu - r }{ f(y, z)}  \right \rangle \, h(z) \, x \, \hp_{0, 0, xz}.
\label{eqn:475} 
\end{equation}
with boundary conditions $\hp_{1,0}(x_{c/r, 0, 0}(z), z) = 0$ and $\hp_{1,0}(x_{0, 0, 0}(z), z) = 0$ at the boundaries $x_{c/r, 0, 0}(z) \equiv 0$ and $x_{0, 0, 0}(z) = \frac{B_1(z) }{ B_1(z) - 1} \cdot \frac{r}{c}$.  It follows that $\hp_{1, 0}$ is given by
\begin{equation}
\hp_{1,0}(z) = x^{B_1(z)} \, \ln \left( x \cdot  \frac{B_1(z) - 1 }{ B_1(z)} \cdot \frac{c}{r} \right) \, \left[ A_1(z) + A_2(z) \ln \left( x \cdot  \frac{B_1(z) }{ B_1(z) -1} \cdot \frac{r}{c} \right) \right] ,
\label{eqn:476} 
\end{equation}
in which $A_1$ and $A_2$ are
\begin{equation}
A_1(z) =  \frac{H_1(z) }{ (2 B_1(z) - 1) \, s(z) - (r - \la)} - \frac{H_2(z) \, s(z) }{ \left[ (2 B_1(z) - 1) \, s(z) - (r - \la) \right]^2},
\label{eqn:477} 
\end{equation}
and
\begin{equation}
A_2(z) =  \, \frac{1}{2}\cdot \frac{H_2(z) }{  (2 B_1(z) - 1) \, s(z) - (r - \la)},
\label{eqn:478} 
\end{equation}
with $H_1$ and $H_2$ functions of $z$ defined by
\begin{equation}
\begin{split}
& H_1(z) + H_2(z) \, \ln x \\
& = - \rho_{13} \, h(z) \left \langle \frac{\mu - r }{ f(y, z)}\right \rangle \frac{B'_1(z) }{ B_1(z) - 1} \left(  
\frac{B_1(z) - 1}{ B_1(z)} \cdot \frac{c}{r} \right)^{B_1(z)} \left[ 1 + B_1(z) \, \ln \left( x \cdot  \frac{B_1(z) - 1 }{ B_1(z)} \cdot \frac{c}{r} \right) \right].
\end{split}
\label{eqn:479} 
\end{equation}

\subsection{The Approximation of the Probability of Lifetime Ruin and the Optimal Investment Strategy}\label{eq:app-tems}
Combining (\ref{eqn:459}), (\ref{eqn:470}), and (\ref{eqn:476}), we obtain the following approximation of $\hp$
\begin{equation}
\begin{split}
\hat{\psi}^{\eps,\delta}(x, z) &= \hp_{0, 0}(x, z) + \sqrt{\eps} \, \hp_{0, 1}(x, z) + \sqrt{\del} \, \hp_{1, 0}(x, z)  \\
&= - \, \frac{1 }{ B_1(z) - 1} \left( \frac{c}{r} \cdot \frac{B_1(z) - 1 }{ B_1(z)} \cdot x  \right)^{B_1(z)} + \frac{c}{r} \, x \\
& \quad + \sqrt{\eps} \, A(z) \, x^{B_1(z)} \,  \ln \left( x \cdot  \frac{B_1(z) - 1 }{ B_1(z)} \cdot \frac{c}{r} \right) \\
& \quad + \sqrt{\del} \, x^{B_1(z)} \, \ln \left( x \cdot  \frac{B_1(z) - 1 }{ B_1(z)} \cdot \frac{c}{r} \right) \, \left[ A_1(z) + A_2(z) \ln \left( x \cdot  \frac{B_1(z) }{ B_1(z) -1} \cdot \frac{r}{c} \right) \right],
\end{split}
\label{eqn:480} 
\end{equation}
in which $A,$ $A_1,$ and $A_2,$ are specified in $(\ref{eqn:469}),$ $(\ref{eqn:477}),$ and $(\ref{eqn:478}),$ respectively.

We also approximate the dual of the optimal investment strategy up to the first powers of $\sqrt{\eps}$ and $\sqrt{\del}$, as we did for $\hp$. Using (\ref{eqn:424}), we obtain
\begin{equation}
\begin{split}
\hat{\pi}^{\epsilon,\delta} (x, z) &= - \frac{\mu - r}{ f^2(y, z)} \, x \, \hp_{0, 0, xx} + \sqrt{\eps} \left( - \frac{\mu - r }{ f^2(y, z)} \, x \, \hp_{0, 1, xx} + \rho_{12} \, \frac{\nu \sqrt{2} }{ f(y, z)} \, \hp_{0, 2, xy}  \right) \\
& \quad + \sqrt{\del} \left( - \frac{\mu - r }{ f^2(y, z)} \, x \, \hp_{1, 0, xx} + \rho_{13} \,  \frac{h(z) }{ f(y, z)} \, \hp_{0, 0, xz} \right).
\end{split}
\label{eqn:51} 
\end{equation}

Given $w \in \mathbb{R}_+$, we solve for $x$ using $w = \hat{\psi}^{\epsilon,\delta}_x(x, z)$. Then, we let $\psi^{\epsilon,\delta}(w,z):=\hat{\psi}^{\eps,\delta}(x, z) - x w$, thereby performing the calculation in equation \eqref{eqn:413}. We also denote by $\pi^{\eps,\delta}$ the function that satisfies $\pi^{\eps,\delta}(w,z):=\hat{\pi}^{\epsilon,\delta} (x, z)$. Note that the resulting approximation of $\psi$ is not guaranteed to be a probability; however, this is not a problem in the numerical experiments we consider in the next section.

\section{Numerical Solution using the Markov Chain Approximation Method}\label{sec:MCAM}

In this section, we describe how to construct a numerical algorithm for the original optimal control problem directly using the Markov Chain Approximation Method (MCAM); see e.g. \citep{KushnerBook, KushnerConsistency}. For the ease of presentation, we will describe the numerical algorithm only when the fast scale volatility factor is present. In what follows $\rho$ will denote the correlation between the Brownian motion driving the stock and the one driving the fast factor, that is, $\rho=\rho_{12}$.

Let us fix an $h$-grid, that is, a rectangular compact domain $G^h  \subset \R^2$ with the same spacing $h$ in both directions. 
We choose an initial guess (on this grid) for a candidate optimal strategy. Denote this strategy by $\pi$. Then, our goal is to create a discrete-time Markov chain $(\xi^h_n)_{n \geq 0}$ that lives on $G^h$ and that satisfies the local consistency condition
\begin{equation}\label{eq:local-const}
\begin{split}
\E_{x,n}^{h,\pi} [\Delta \xi_{n+1}^h ]&= b(x,\pi)\Delta t^{\pi,h}(x,\pi) + o(\Delta t^{h}),\\
\text{Cov}_{x,n}^{h,\pi} [\Delta \xi_{n+1}^h ]&= A(x,\pi)\Delta t^{\pi,h}(x,\pi) + o(\Delta t^{h}),
\end{split}
\end{equation}
in which $\Delta \xi_{n+1}=\xi_{n+1}-\xi_n$, and $b$ and $A$ denote the drift and the covariance of the vector $X_t=(W_t,Y_t)$, respectively. (The Markov chain is constructed to approximate this vector in a certain sense.) 
$\mathbb{E}_{x,n}^{h,\pi}$ denotes the expectation, given that the state of the Markov chain at time $n$ is $x$. In \eqref{eq:local-const} the quantity $\Delta t^{h}$ (called the interpolation interval) is to be chosen so that it goes to zero as $h \to 0$. We also do not want this quantity to depend on the state variables or the control variable. 

 Since $G^h$ is a compact domain, we impose reflecting boundary conditions at its edges. (Natural boundaries exist for $W(t)$, specifically  $0$ and the safe level $\frac{c}{r}$. However, $Y_t$ lives on an infinite region.) For example, we choose the transition probabilities to be $p^{\pi,h}((w,y),(w,y-h))=1$, when $y$ is as large it can be in $G^h$ and for all $w \in [0, \frac{c}{r}]$.
 
\subsection{Constructing the Approximating Markov Chain}\hfill

\subsubsection{When $\rho=0$.} Denote $\alpha=\frac{1}{\epsilon}, \beta=\nu \sqrt{\frac{2}{\epsilon}}$. We obtain the transition probabilities of the Markov chain $\xi^h$ as 

\begin{equation}
\begin{cases}
      p^{\pi,h}((w,y), (w,y \pm h))& =\displaystyle \frac{ \beta^2/2 + h \alpha (m-y)^{\pm} }{\widetilde{Q}^h}\,, \\ 
       p^{\pi,h}((w,y), (w\pm h ,y ))& =\displaystyle  \frac{(f(y)\pi(w,y))^2/2 + h (\mu-r) \pi(w,y)^{\pm}+  h (rw-c)^{\pm}}{\widetilde{Q}^h}\, ,\\
         p^{\pi,h}((w,y), (w,y ))&= \frac{\widetilde{Q}^h-Q^{\pi,h}(w,y)}{\widetilde{Q}^h}\,,
       \end{cases}
\end{equation}   
and choose the interpolation interval to be
$$\Delta t^{h} =\frac{h^2}{ \widetilde{Q}^h},$$
in which $$Q^{\pi,h}(w,y)= (\pi f(y))^2 + \beta^2 + h |\alpha (m-y)| + h|(\mu-r)\pi(w,y)| + h|rw-c| ,$$ and
\[
\widetilde{Q}^h=\max_{(w,y, \pi)}  Q^{\pi,h}(w,y),
\]
in order to satisfy the local consistency condition. Here $a^{\pm}=\max\{0, \pm a\}$.
\subsubsection{When $\rho \neq 0$.} In this case a convenient transition probability matrix solving the local consistency condition is
\begin{equation} \label{eq:tranprob}
\begin{cases}
      & p^{\pi,h}((w,y), (w,y \pm h)) =\displaystyle \frac{(1-\rho^2)\beta^2/2 -|\rho  \pi(w,y)| \beta f(y)/2 + h \alpha (m-y)^{\pm} }{\widetilde{Q}^h}\,, \\ 
      & p^{\pi,h}((w,y), (w\pm h ,y )) =\displaystyle  \frac{(f(y)\pi(w,y))^2/2-|\rho \pi(w,y)| \beta  f(y)/2  + h (\mu-r) \pi(w,y)^{\pm}+  h (rw-c)^{\pm}}{\widetilde{Q}^h}\,,\\
       & p^{\pi,h}((w,y), (w+ h ,y+h )) =  p^{\pi,h}((w,y), (w-h ,y-h )) =\displaystyle  \frac{ (\rho \pi(w,y))^{+}  \beta f(y)}{2\widetilde{Q}^h}, \;  \\
        &  p^{\pi,h}((w,y), (w+ h ,y-h )) =  p^{\pi,h}((w,y), (w-h ,y+h )) = \displaystyle \frac{( \rho \pi(w,y))^{-} \beta  f(y)}{2\widetilde{Q}^h}\,\\
        & p^{\pi,h}((w,y), (w,y ))= \frac{\widetilde{Q}^h-Q^{\pi,h}(w,y)}{\widetilde{Q}^h}\,,
       \end{cases}
\end{equation}   
where $$Q^{\pi,h}(w,y)=  (\pi f(y))^2 + \beta^2 -|\rho \pi(w,y)| \beta  f(y) + h |\alpha (m-y)| + h|(\mu-r)\pi(w,y)| + h|rw-c| .$$

For values of $|\rho|$ close to 1, the transition probabilities may be negative. The positiveness of these probabilities is equivalent to the diagonal dominance of the covariance matrix $A=(a_{ij})$. (Recall that we call $A$ diagonally dominant if $
a_{ii}-\sum_{j,j\neq i}|a_{ij}| > 0, \quad \forall i$.) The construction of an approximating Markov chain when some of the expressions in \eqref{eq:tranprob} are negative will be discussed next. 

\subsubsection{When $\rho=1$ and some of the transition probabilities in \eqref{eq:tranprob} are negative.} We accomplish the construction of the approximating Markov chain in two steps, following \citet{KushnerConsistency}:

\noindent \textbf{(i) Decomposition.} 
As in \citet{KushnerBook} Sections 5.3 and 5.4, we decompose $X$ into separate components and build approximating Markov chains to match each component. Then, we combine the transition probabilities appropriately to obtain the approximating Markov chain for $X$ itself.

Let $X=X^{(1)}+X^{(2)}$, in which
\begin{eqnarray}
\dd X^{(1)}_t &=& \begin{pmatrix} \pi f(y)   \\ \beta  \end{pmatrix} \dd B^1_t ,\label{eqn:dx1}\\
\dd X^{(2)}_t &=& \begin{pmatrix} rW_t-c + (\mu - r)\pi_t \\ \alpha (m-Y_t) \label{eqn:dx2} \end{pmatrix} dt.
\end{eqnarray}
Since $\rho=1$, we take $B^1=B^2$.
Suppose that the form of the locally consistent (with dynamics of $X^{(1)}$ and $X^{(2)}$, respectively) transition probabilities and interpolation intervals are
\begin{eqnarray*}
 p_1^{\pi,h}(x,\bar{x}) &= &\frac{n_1^{\pi,h}(x,\bar{x})}{\widetilde{Q}_1^{h}}\; , \;\Delta t_1^{\pi,h} =\frac{h^2}{\widetilde{Q}_1^{h}}\; , \\
  p_2^{\pi,h}(x,\bar{x}) &= & \frac{n_2^{\pi,h}(x,\bar{x})}{\widetilde{Q}_2^{h}}\; , \; \Delta t_2^{\pi,h} =\frac{h}{\widetilde{Q}_2^{h}}\;,
  \end{eqnarray*}
  for some $n_1^{\pi,h}(x,\bar{x}), n_2^{\pi,h}(x,\bar{x})$, and appropriate normalizers $\widetilde{Q}_1^{h}$, $\widetilde{Q}_2^{h}$. 
Then, the following transition probabilities and the interpolation interval are locally consistent with the dynamics of $X$ 
\begin{eqnarray}
p^{\pi,h}(x,\bar{x})  =\displaystyle \frac{n_1^{\pi,h}(x,\bar{x})  + h n_2^{\pi,h}(x,\bar{x}) }{\widetilde{Q}_1^{h}+h \widetilde{Q}_2^{h}}, \quad \Delta t^{\pi,h} =\displaystyle\frac{h^2}{ \widetilde{Q}_1^{h}+h \widetilde{Q}_2^{h}}.
\label{eqn:pdx}
\end{eqnarray} 

Since it is easier, we first provide the  expression for $p^{\pi,h}_2$:
\begin{equation}
\begin{cases}
    p^{\pi,h}_2((w,y), (w,y \pm h) | \pi)& =\displaystyle \frac{   \alpha (m-y)^{\pm} }{\widetilde{Q}_2^{h}}\,, \\
           p^{\pi,h}_2((w,y), (w\pm h ,y ) | \pi)& =\displaystyle  \frac{  (\mu-r) \pi(w,y)^{\pm}+   (rw-c)^{\pm}}{\widetilde{Q}_2^{h}}\,,
           \\ p_2^{\pi,h}((w,y), (w,y ))&= \displaystyle \frac{\widetilde{Q_2}^h-Q_2^{\pi,h}(w,y)}{\widetilde{Q}_2^h},
       \label{eqn:pdx2}
\end{cases}
\end{equation}
 where
 $$Q_2^{\pi,h}(w,y)= \displaystyle \alpha |m-y| +  (\mu-r) |\pi(w,y)|+   |rw-c| .$$

The computation of $p^{\pi,h}_1$ is more involved. This is the subject of the next step.  
 
\noindent \textbf {(ii) Variance control.} System (\ref{eqn:dx1}) is fully degenerate; that is, the corresponding covariance matrix $A$ is not diagonally dominant.  The previous technique for building a Markov chain does not work. Instead, we will build an approximating Markov chain by allowing the local consistency condition to be violated by a small margin of error.

If $(\sigma_1,\sigma_2) =  (q k_1,q k_2)$ for some constant $q$ and integers $k_1$, $k_2$, we could let the transition probability to be $p^{h}(x,x \pm (h k_1,h k_2)) = {1}/{2}$ and the interpolation interval to be $\Delta t^{h} = {h^2}/{q^2}$, and we would obtain a locally consistent Markov chain. This is not possible in general. For an arbitrary vector $(\sigma_1,\sigma_2)$, 
we can find a pair of integers $k_1(x,\pi), k_2(x,\pi)$, and a real number $\gamma(x,\pi) \in [0,1]$, such that 
$$\begin{pmatrix}   \sigma_1(x,\pi)  \\ \sigma_2 \end{pmatrix} = q(x,\pi)\begin{pmatrix} k_1(x,\pi)  \\k_2(x,\pi) + \gamma(x,\pi) \end{pmatrix}. $$

Since the Markov chain is constrained to the grid $G^h$, we can only approximately let it move in the direction of $(\sigma_1,\sigma_2)^T$. We choose
 \begin{equation}
 \begin{split}
 p^{\pi,h}(x,x\pm h(k_1, k_2 )^T) & = p_1/2 , \\
  p^{\pi,h}(x,x\pm h(k_1, k_2+1)^T) & = p_2/2 , \label{eqn:pdx1}
 \end{split}
\end{equation}
in which $p_1 + p_2 =1$, and $p_1$ and $p_2$ will be appropriately chosen in what follows. 
The mean and the covariance of the approximating chain is  
\begin{equation}
\begin{split}
\E_{x,\pi}^{h,\pi}[\Delta \xi^h(x,\pi)] &= 0,\\
\E_{x,\pi}^{h,\pi}[\Delta \xi^h(x,\pi)\Delta \xi^h(x,\pi)^T] &= h^2 C(x,\pi),
\end{split}
\end{equation}
where
\begin{equation}
\begin{split}
C(x,\pi)&= p_1 \begin{pmatrix} k_1^2 & k_1 k_2 \\ k_1 k_2 & k_2^2\end{pmatrix} + p_2 \begin{pmatrix}k_1^2 & k_1 (k_2 +1)\\ k_1 (k_2+1) & (k_2+1)^2 \end{pmatrix} \\
 & = \begin{pmatrix} k_1^2 & k_1 (k_2+p_2) \\ k_1 (k_2+p_2) & k_2^2 + 2 p k_2 + p_2 \end{pmatrix}.
 \end{split}
 \label{eqn:ccc}
\end{equation}

We choose the interpolation interval to be $ \Delta t^{\pi,h}(x,\pi)=h^2/q^2$. 
On the other hand
$$a(x,\pi)= A(x,\pi)/q^2=  \begin{pmatrix} k_1^2 & k_1 (k_2+\gamma) \\ k_1 (k_2+\gamma) & (k_2+\gamma)^2  \end{pmatrix},$$
and we see that if we pick $p_2 = \gamma $,
then $C_{11}=a_{11}$ and $C_{12}=a_{12}$ match, but we violate the local consistency condition 
by
\begin{equation}\label{eq:var-cont-noise}
\frac{C_{22} - a_{22}^2}{a_{22}^2} = \frac{\gamma (1-\gamma)}{ (k_2 + \gamma)^2} = O\left(\frac{1}{k_2^2}\right).
\end{equation}
We will choose $k_2$ sufficiently large so that the local consistency condition is almost satisfied, and the numerical noise in \eqref{eq:var-cont-noise} is significantly reduced.

\subsubsection{The case when $\rho \in (-1,1)$ and some of the transition probabilities in \eqref{eq:tranprob} are negative} 
We will decompose the state variable into three components:
\begin{equation}
\dd \vec{X_t} = \begin{pmatrix} \dd W_t \\ \dd Y_t \end{pmatrix} = \begin{pmatrix}rW_t-c + (\mu - r)\pi_t \\ \alpha (m-Y_t)\end{pmatrix} dt + \begin{pmatrix}\pi_t f(Y_t)  \\ 
\beta \rho  \end{pmatrix} \dd B_t^1  + \begin{pmatrix}0 & 0 \\ 
0 & \beta \sqrt{1-\rho^2} \end{pmatrix} \begin{pmatrix} \dd B_t^1 \\ \dd B_t^2 \end{pmatrix}, 
\label{eqn:dx3}
\end{equation}
 that is, a drift component, a fully degenerate noise component, and a noise component with diagonally dominated covariance matrix. We can build an approximating Markov chain for each component separately and 
 then combine them as discussed above.
 
\subsection{Approximating the Probability of Ruin and Updating the Strategy} 
 
 We solve the system of linear equations
 \begin{equation}\label{eq:dy-pg-pr}
 V^{\pi,h}(x)=e^{-\lambda \Delta t^{\pi,h}}\sum_{\tilde{x} \in G^h}p^{\pi,h}(x,\tilde{x})V^{\pi,h}(\tilde{x}),
 \end{equation}
 with boundary conditions $V^{\pi,h}(0,y)=1$ and $V^{\pi,h}(c/r,y)=0$.
 This is the dynamic programming equation for a probability of ruin problem when the underlying state variable is the Markov chain $\xi^h$. In the next step, we update our candidate for the optimal strategy. 
For convenience, denote $V^{\pi,h}$ by $V$.
 In the interior points of the grid
{\small
 \[
 \pi(w,y)=-\frac{h(\mu-r)[V(w+h,y)-V(w,y)]+ (\beta/2) \rho f(y)\left[V(w+h,y+h)+V(w,y-h)-V(w+h,y-h)-V(w,y+h)\right]}{f^2(y) \left[V(w+h,y)+V(w-h,y)-2V(w,y)\right]}.
 \]
} 
 On the wealth dimension boundaries of the grid, we let  $\pi(c/r,y)=0$ and 
 \[
 \begin{split}
\pi(0,y)=-\frac{h(\mu-r)[V(h,y)-V(0,y)]+(\beta/2) \rho f(y)\left[V(h,y+h)+V(0,y-h)-V(h,y-h)-V(0,y+h)\right]}{f^2(y) \left[2V(0,y)-5V(h,y)+4V(2h,y)-V(3h,y)\right]}.
 \end{split}
 \]
 The updates of the optimal strategy for the maximum and minimum values of $y$ are similar.
 
 \subsection*{Iteration}
Once the optimal strategy is updated, we go back and update the transition probabilities and solve the system of linear equations in \eqref{eq:dy-pg-pr} to update the value function. This iteration continues until the improvement in the value function is smaller than an exogenously picked threshold.
 
 \subsection*{Two Technical Issues}

\begin{itemize}
	\item The initial guess of the optimal strategy is important. For $\rho=0$, we take the initial strategy as the one in constant volatility case, where the closed-form solution is available in \citet{young}. For $\rho \neq  0$, we take the final strategy computed from zero-correlation case ($\rho=0$) as the initial guess. This initial guess makes the algorithm converge fast.
	
	\item For $\rho \neq 0$, the covariance matrix of the wealth process and volatility factor, in general, does not satisfy the diagonal dominance condition. The problem is more serious for the slow factor, since its variance is of the order of $\delta$, and the numerical noise using "`variance control"' is far greater.
	To solve this issue we perform a ``scale adjustment" to increase the variance of the factor. 	
	For example,  if  we define $\bar{Z_t}=100 Z_t$, then the dynamic of the system becomes
	\begin{equation}
	\begin{split}
	\frac{\dd S_t}{S_t} &= \mu dt + f\left(Y_t,\frac{\bar{Z_t}}{100}\right) \dd B_t^1,\\
	\dd \bar{Z_t}&= \delta (100 m-\bar{Z_t}) dt + 100 \sqrt{\delta} \sqrt{2} \nu_2 \dd B_t^{(3)}. 
	\end{split}
	\end{equation}
	when $g=(m-z)$ and $h=\sqrt{2} \nu$.
	The new system is mathematically equivalent to the original one, but with a much bigger variance; thus, the numerical noise in variance control is much smaller. 	Note that the scheme here is equivalent to choosing a different grid sizes for the volatility and wealth dimensions.
\end{itemize}

\section{Numerical Experiments}

In order to conduct our numerical experiments we will take the dynamics of the slow factor in \eqref{eqn:27} to be 
\begin{equation*}
\dd Z_t = \delta (m-Z_t)dt + \sqrt{\delta} \sqrt{2}\nu \dd B_t^{(3)},\  \quad Z_0=z.
\end{equation*}
We let $f(y,z)=\exp(-y)$ or $f(y,z)=\exp(-z)$ in \eqref{eqn:sigmat}, 
depending on whether we want to account for the fast volatility factor or the slow volatility factor in our modeling. 
We will call $1/\eps$ or $\delta$ the speed of mean reversion. We will take the correlations between the Brownian motions driving the volatility factors and the stock price to be $\rho=\rho_{13}=\rho_{12}$.

The following parameters are fixed throughout this section:
\begin{itemize}
	\item $r=0.02$; the risk-free interest rate is 2\% over inflation.
	\item $\mu=0.1$; the expected return of risky asset is 10\% over inflation.
	\item $c=0.1$; the individual consumes at a constant rate of 0.1 unit of wealth per year. 
	\item $\lambda=0.04$; the hazard rate (force of mortality) is constant such that the expected future lifetime is always 25 years.
	\item $m=1.364$ and $\nu=0.15$, so that the harmonic average volatility, which we will denote by $\sigma_m = \sqrt{1/\E[1/f^2(Y)]} = \sqrt{1/\E[e^{2Y}]}=e^{-m-\nu^2}=0.25$, in which $Y$ is a normal random variable with mean $m$ and variance $\nu^2$. The distribution of this random variable is the stationary distribution of the process $(Y_t)_{t \geq 0}$; see \eqref{eqn:446}. [Note that $\sigma_m$ is very close in value to $\E[f(Y)]=\E[e^{-Y}]=e^{-m+\nu^2/2}=0.26$.]
	\end{itemize}
In our numerical procedure we use a bounded region for $Y$ and impose reflecting boundary conditions. 
However, $f(Y_t)$ is not bounded and not bounded away from zero. 
On the other hand, the invariant distribution of the process $Y$ is normal with mean 1.364, and variance $0.15^2$. So it is with very small probability that $Y_t$ is negative or very large. Therefore, the fact that $f(Y_t)$ is not bounded or bounded away from zero does not affect the accuracy of our numerical work in a significant way.
	
\subsection*{Observation 1}	
	
\begin{figure}
  \centering
   \subfloat[Minimum probability of ruin ]{\label{fig:mcam3a} \includegraphics[width=0.75\textwidth]{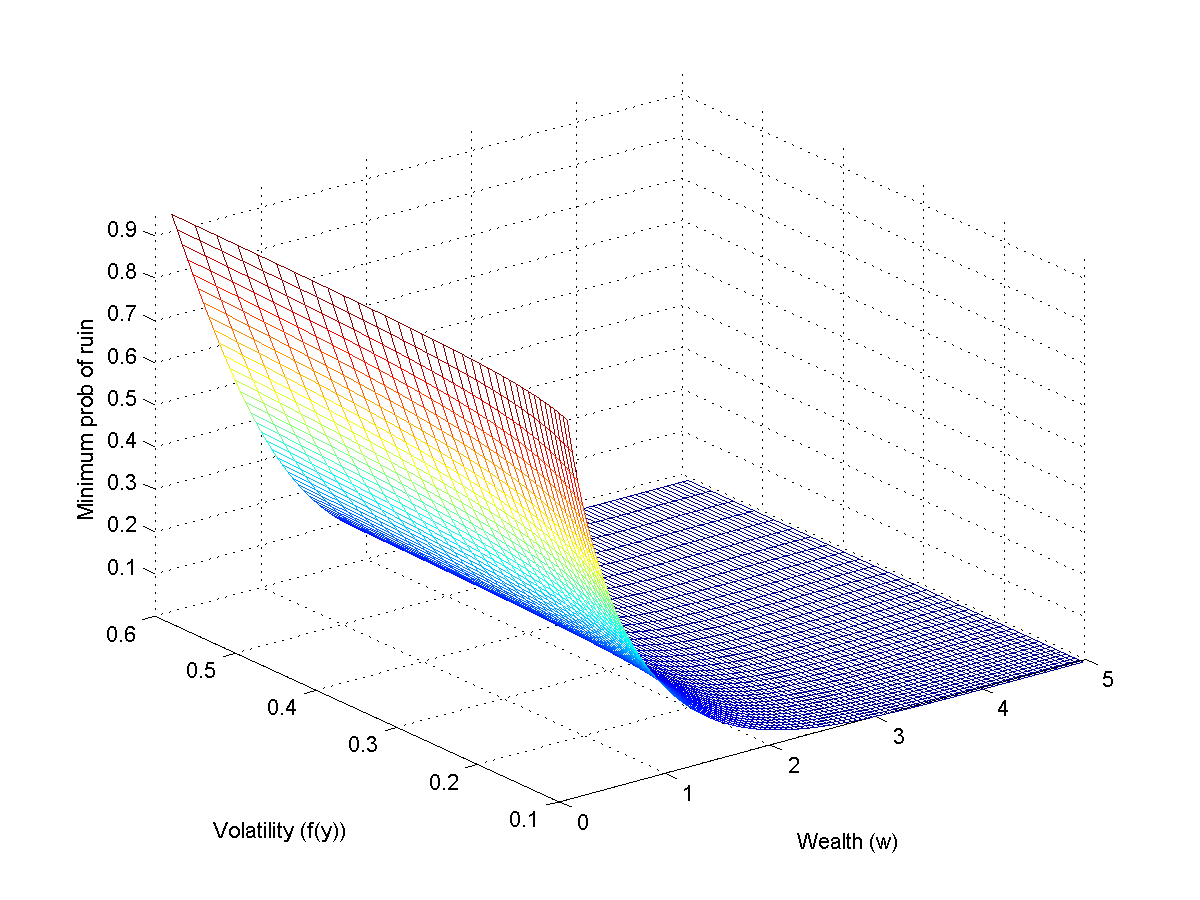}}    \\           
  \subfloat[Optimal investment strategy ]{\label{fig:mcam3b}\includegraphics[width=0.75\textwidth]{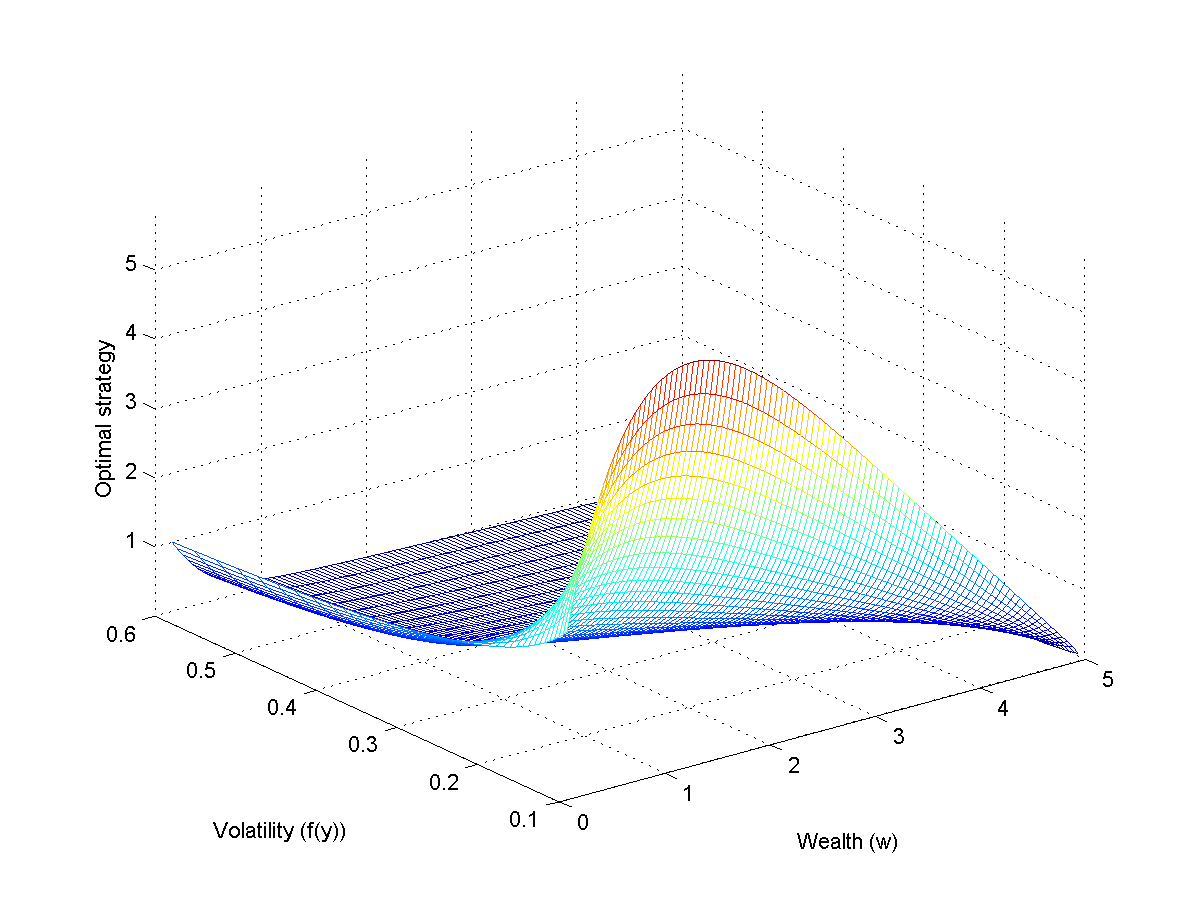}}
  \caption{Minimum probability of ruin and optimal strategy computed by MCAM. Speed of mean reversion$=0.5$.}
  \label{fig:mcam3}
\end{figure}

We give a three-dimensional graph of the minimum probability of ruin and the optimal investment strategy in Figure~\ref{fig:mcam3}, which are computed using MCAM. Here the speed of mean reversion is 0.5, $\rho=0$, and only one factor is used. In our experiments we observed that the optimal strategy $\pi^*$ is positive (no-shortselling).
As expected we observe that $w \to \psi(w,y)$ is convex and decreasing. Note that $f(y) \to \psi(w,y)$ is increasing. Also, $f(y) \to \pi^*(w,y)$ is decreasing; however, it is not necessarily true that $w \to \pi^{*}(w,y)$ is decreasing. The latter behavior depends on the value of $y$.

\begin{figure}
  \centering   
  \subfloat[Probability of ruin at $\sigma_0$=0.6 ]{\includegraphics[width=0.49\textwidth]{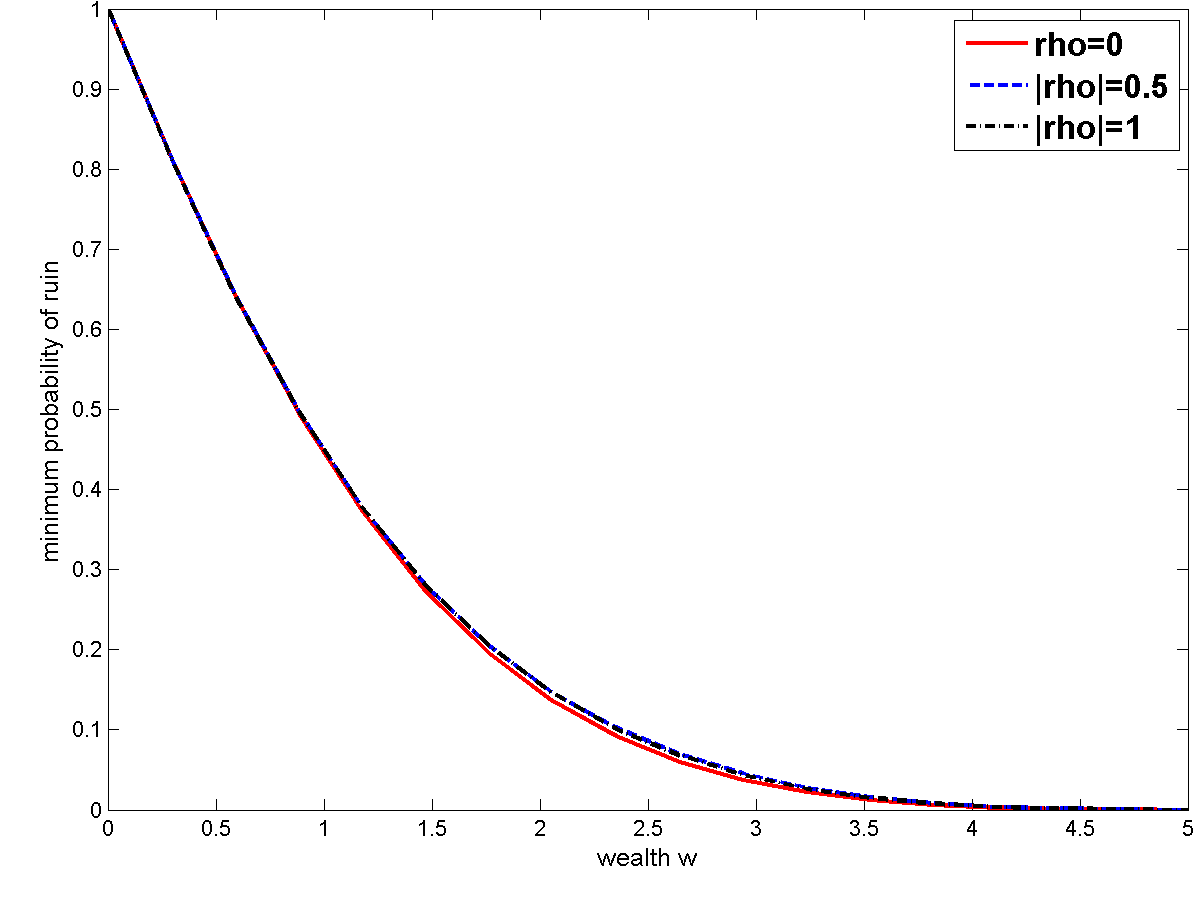}}    \\           
  \subfloat[Probability of ruin at $\sigma_0$= $\sigma_m$=0.25 ]{\includegraphics[width=0.49\textwidth]{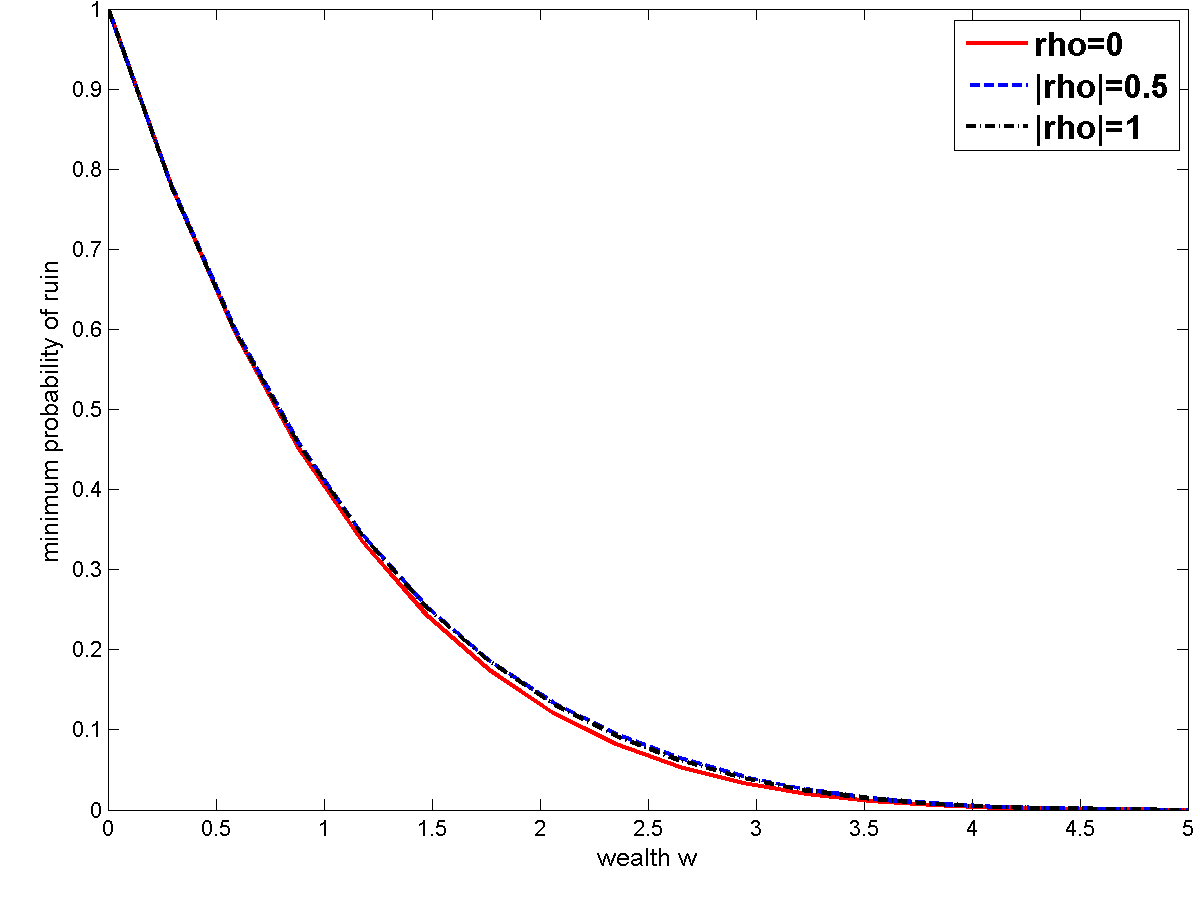}} \\
  \subfloat[Probability of ruin at $\sigma_0$=0.1 ]{\includegraphics[width=0.49\textwidth]{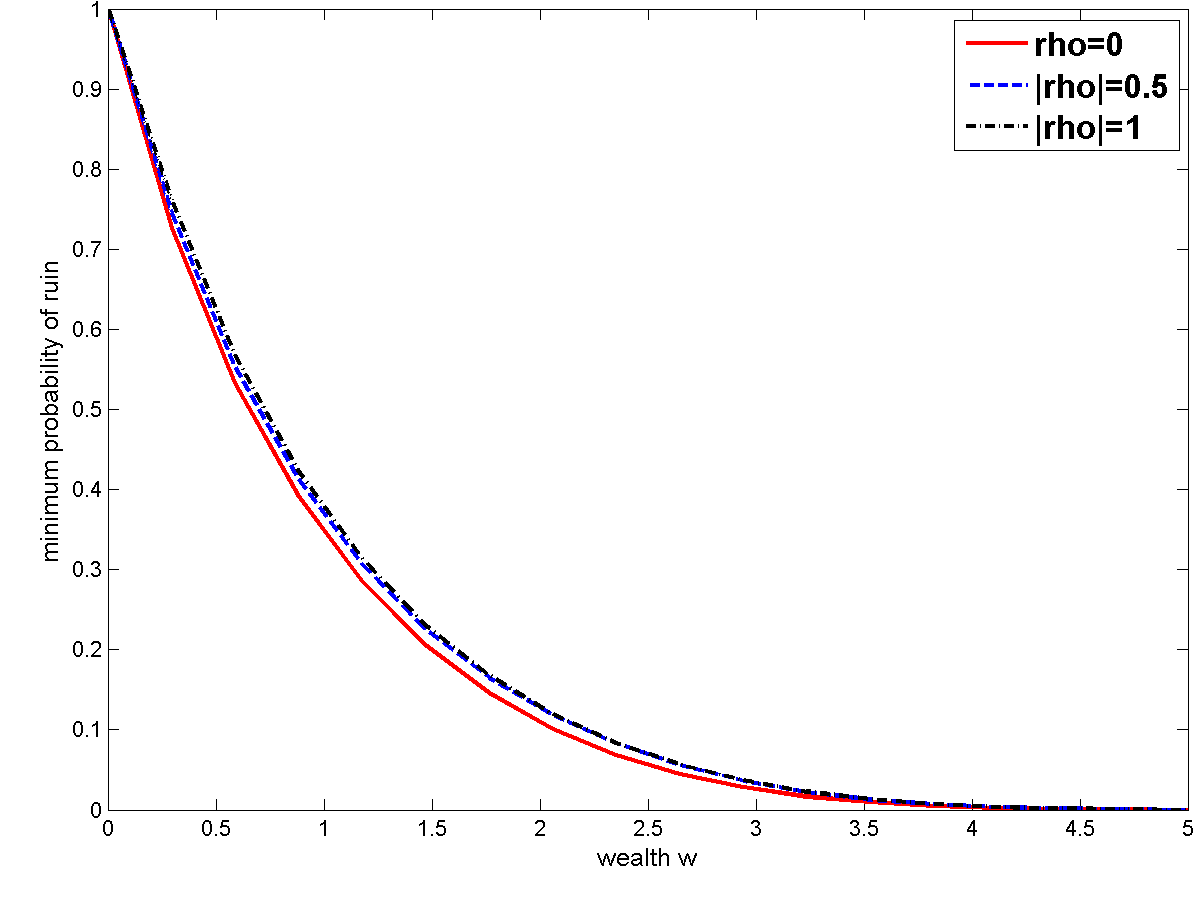}} 
  \caption{Variations of the minimum probability of ruin with respect to $\rho$. Speed of mean reversion$=0.5$. }
  \label{fig:rho3}
\end{figure}

The probability of ruin does not depend on the sign of the correlation, $\rho$, between the Brownian motions driving the stock and the one driving the volatility. The larger the magnitude of $\rho$, the larger the probability of ruin. However, the minimum probability of ruin is quite insensitive to the changes in $\rho$; see Figure~\ref{fig:rho3}.

\subsection*{Observation 2}
We compare the optimal investment strategy $\pi^*(w,y,z)$ in \eqref{eqn:414} to 
\begin{equation*} 
\widetilde{\pi}(w;\sigma)=\frac{\mu-r}{\sigma^2} \frac{c-rw}{(p-1)r}\;\;,
\end{equation*}
in which
\[
p=\frac{1}{2r}\left[(r+\lambda+s)+\sqrt{(r+\lambda+s)^2- 4 r \lambda}\right],
\]
and
\[
s=\frac{1}{2}\left(\frac{\mu-r}{\sigma}\right)^2.
\]
When we want to emphasize the dependence on $\sigma$, we will refer to $p$ as $p(\sigma)$.
\citet{young} showed that the strategy $\widetilde{\pi}$ is optimal when the volatility is fixed to be $\sigma$.

\begin{figure}
  \centering
   \subfloat[Minimum probability of ruin ]{\includegraphics[width=0.75\textwidth]{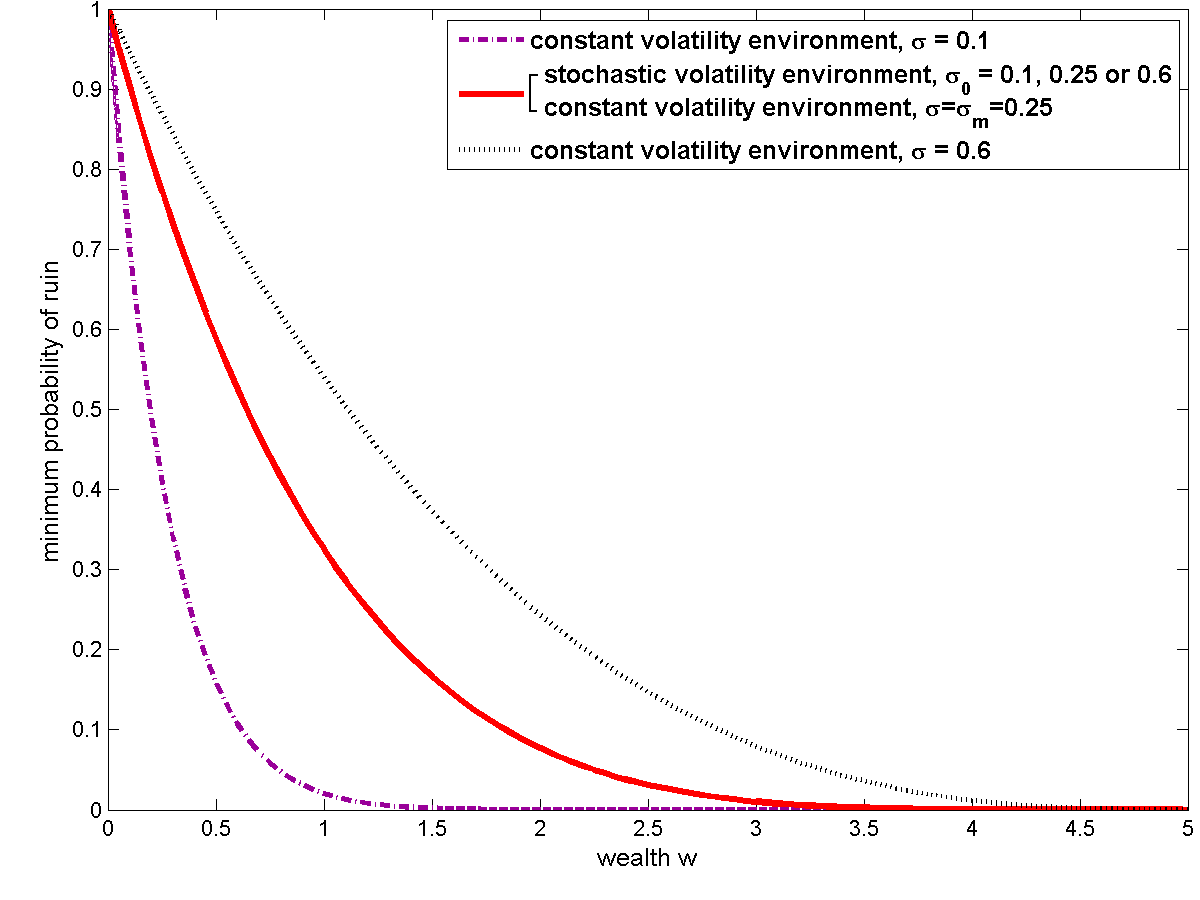}}    \\           
  \subfloat[Optimal investment strategy ]{\includegraphics[width=0.75\textwidth]{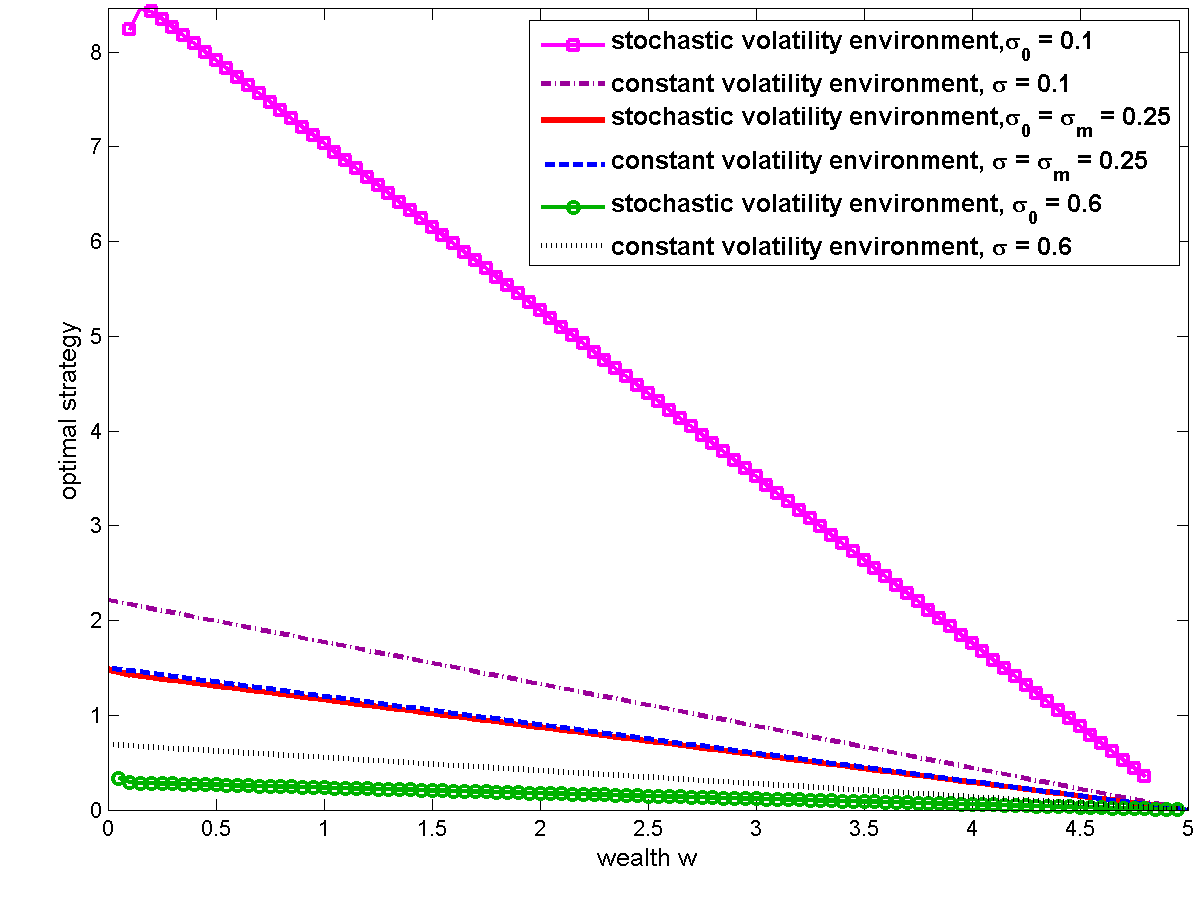}}
  \caption{Stochastic volatility versus constant volatility environment. Speed of mean reversion=$250$.}
  \label{fig:sWorld1}
\end{figure}

If only the fast factor is present and the speed of mean reversion is 250 ($\epsilon=0.004$), then $\hat{\psi}^{\eps,\delta}$ in \eqref{eqn:480} can be expressed as
\[
\hat{\psi}^{\eps}(w)=\hat{\phi}_{0,0}(x)
\]
whose inverse Legendre transform is 
\begin{equation}\label{eq:vgax}
\psi^{\eps}(w;\sigma_m)=\left(1-\frac{r}{c} w\right)^{p(\sigma_m)},
\end{equation}
which is exactly the minimal probability of ruin  if the volatility were  fixed at $\sigma_m$. Therefore, it is not surprising that for very small values of $\eps$, the minimum probability of ruin $\psi(w,y)$, calculated using MCAM, can be approximated by \eqref{eq:vgax}; see Figure~\ref{fig:sWorld1}-a. 
In our numerical calculations and in \eqref{eq:vgax}, 
we observe that the minimum probability of ruin $\psi$ does not depend on its second variable. This result is intuitive, since when only the fast factor is present whatever the initial value of $\sigma_0$ is, the volatility quickly approaches its equilibrium distribution (which is normal with mean $\sigma_m$).
In fact $\pi_{0}(w; \sigma_m)$ practically coincides with the optimal investment strategy $\pi^*$, which is computed using MCAM; see Figure~\ref{fig:sWorld1}-b.

The most important conclusion from Figure~\ref{fig:sWorld1}-b is that it is not necessarily true that the optimal investment strategy when there is stochastic volatility is more or less than the optimal investment strategy when the volatility is constant. Comparing $\widetilde{\pi}(w;\sigma)$ and $\pi^{*}(w,-\ln(\sigma))$ for different values of $\sigma$, we see that $\pi^{*}(w,-\ln(\sigma))<\widetilde{\pi}(w;\sigma)$ for larger values of $\sigma$, whereas the opposite inequality holds for smaller values of $\sigma$. The investment amount decreases significantly as the volatility increases.

\begin{figure}
  \centering
   \subfloat[Minimum probability of ruin ]{\includegraphics[width=0.75\textwidth]{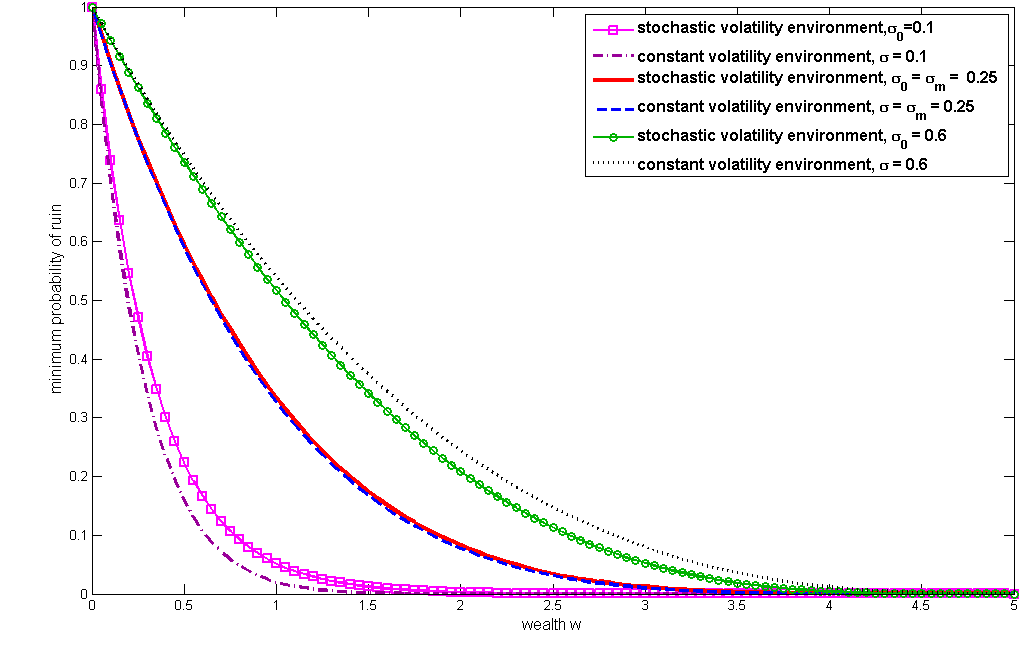}}    \\           
  \subfloat[Optimal investment strategy  ]{\includegraphics[width=0.75\textwidth]{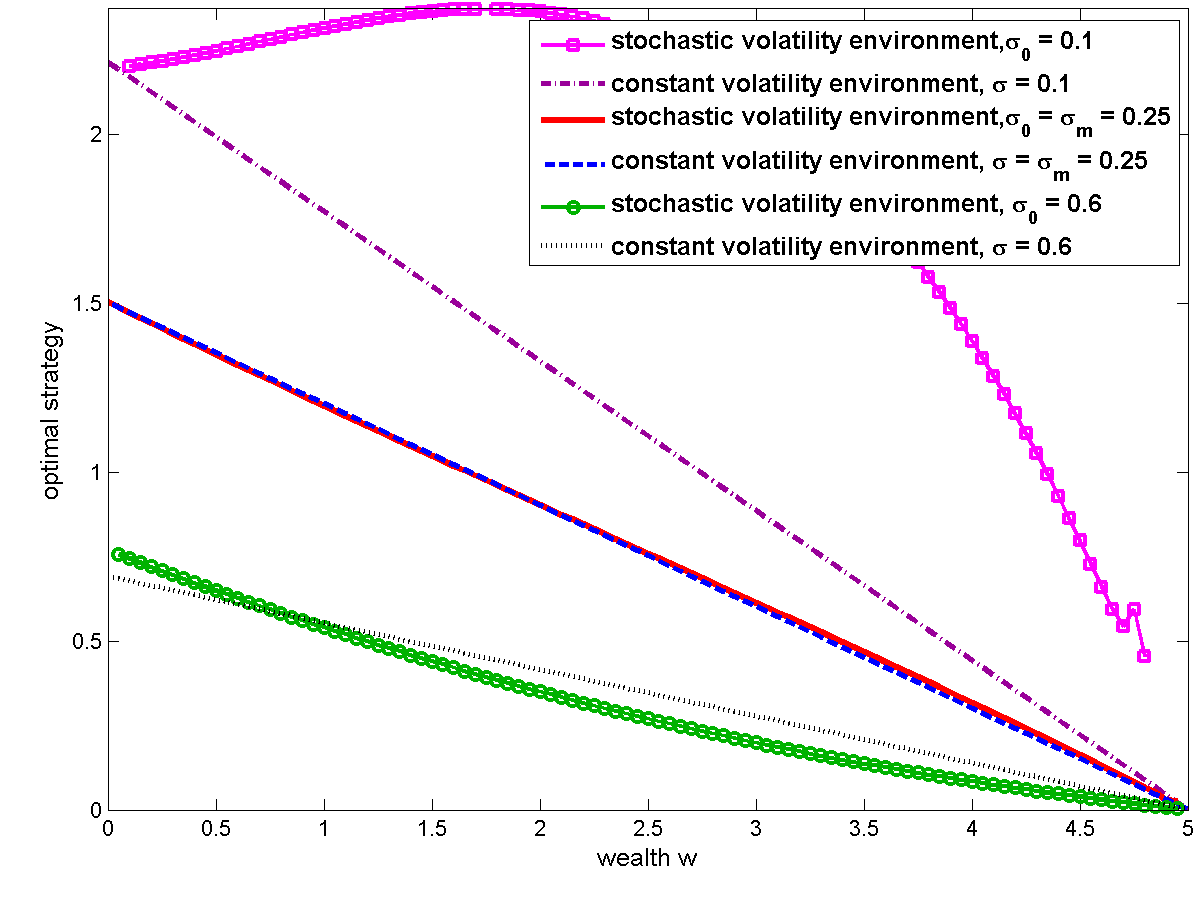}}
  \caption{Stochastic volatility versus constant volatility environment. Speed of mean reversion=$0.02$.}
  \label{fig:sWorld16}
\end{figure}

If only the slow factor is present and the speed of mean reversion is 0.02, then 
\begin{equation}\label{eq:vgaxd}
\psi^{\delta}(w,z)=\left(1-\frac{r}{c} w\right)^{p(e^{-z})},
\end{equation}
approximates the minimum probability of ruin $\psi(w,z)$, which we calculate using MCAM, quite well; compare $\psi(w,-\text{ln}(\sigma))$ and $\psi^{\delta}(w,-\text{ln}(\sigma))$ for different values of $\sigma$ in Figure~\ref{fig:sWorld16}-a. We also compare $\widetilde{\pi}(w;\sigma)$ and $\pi^{*}(w,-\text{ln}(\sigma))$ for several values of $\sigma$ and draw the same conclusions as before. Also note that the optimal investment strategy is not necessarily a decreasing function of wealth.

\begin{figure}
  \centering
   \subfloat[Minimum probability of ruin ]{\includegraphics[width=0.75\textwidth]{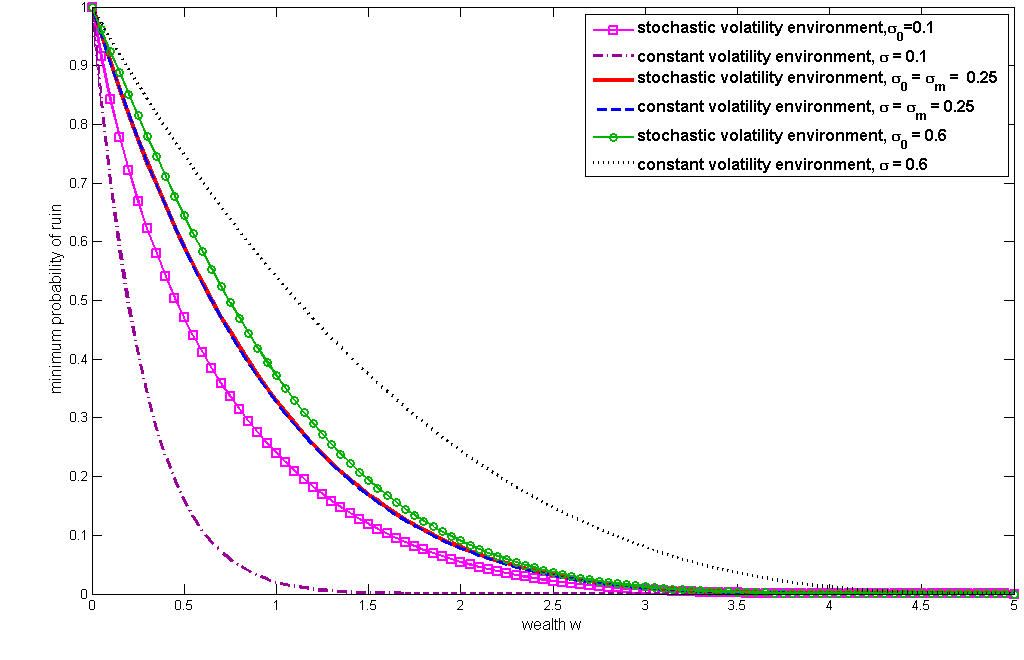}}    \\           
  \subfloat[Optimal investment strategy  ]{\includegraphics[width=0.75\textwidth]{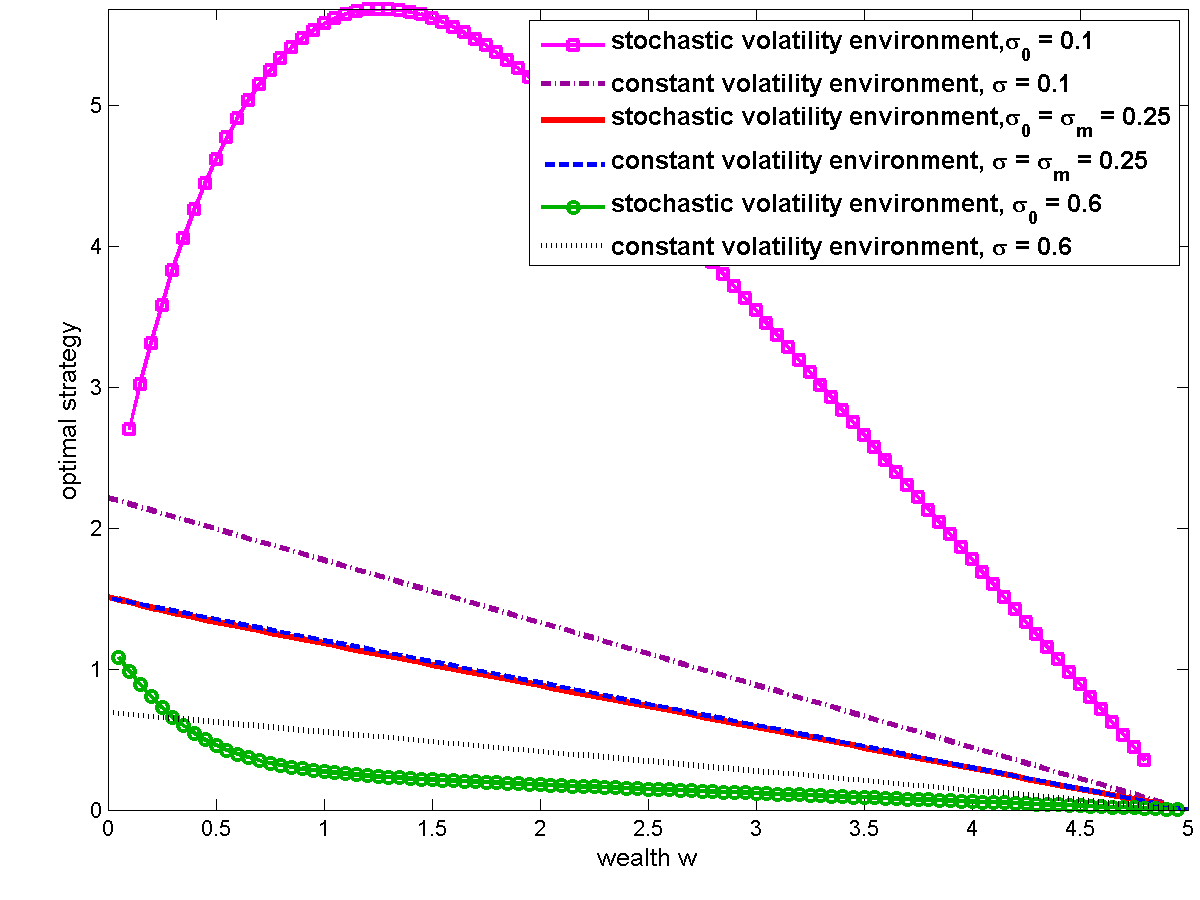}}
  \caption{Stochastic volatility versus constant volatility environment. Speed of mean reversion=$0.5$.}
  \label{fig:sWorld13}
\end{figure}

When we take the speed of mean reversion to be 0.2 (medium speed), then the probability of ruin starts diverting from what \eqref{eq:vgax} or \eqref{eq:vgaxd} describes; see Figure~\ref{fig:sWorld13}-a. As to the comparison of the optimal investment strategy with $\widetilde{\pi}(w;\sigma)$, the same conclusions can be drawn;  see Figure~\ref{fig:sWorld13}-b.

\subsection*{Observation 3}

We compare the performance of several investment strategies in the stochastic volatility environment.
Let $\sigma_0$ be the initial volatility. We denote by $\pi^M$ the strategy when one only invests in the money market. The corresponding probability of ruin can be explicitly computed as $\psi^{M}(w)=(1-c/r w)^{1 \vee [\lambda/r]}$. We will also denote
$\pi^a(w)=\widetilde{\pi}(w;\sigma_0)$, $\pi^{b}=\widetilde{\pi}(w;\sigma_m)$, and
\begin{equation}\label{eq:pi-c}
\pi^{c}(w,y,z)=\frac{\mu-r}{f^2(y,z)} \frac{c-rw}{(p-1)r}.
\end{equation}
Let $\pi^{\eps}(w)$ denote the approximation to the optimal strategy we obtained in Section ~\ref{eq:app-tems} when we only use the $\eps$-perturbation. Similarly, let $\pi^{\delta}(w,z)$ be the approximation to the optimal strategy when we only use the $\delta$-perturbation.

We obtain the probability of ruin corresponding to a given strategy $\pi$ by solving the linear partial differential equation $\D^{\pi} v=0$, (see \eqref{eqn:31} for the definition of the differential operator $\D^{\pi} $) with boundary conditions $v(0,y,z)=0$ and $v(c/r,y,z)=1$. (This computation uses the MCAM without iterating.)

\begin{figure}
  \centering
   \includegraphics[width=0.9\textwidth]{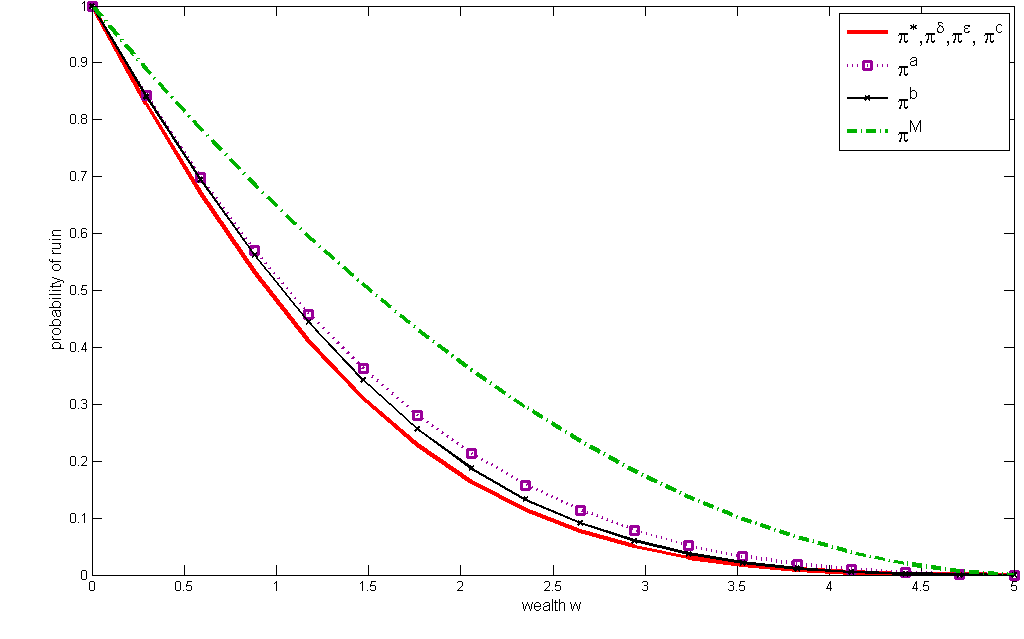}
    \caption{Performance of the investment strategies described in Observation 3 (of Section 6).  Speed of mean reversion=$0.2$. Correlation $\rho=0.5$. Initial volatility $\sigma_0=0.6$. }
  \label{fig:diffStrat109}
\end{figure}

\begin{figure}
  \centering
 \includegraphics[width=0.9\textwidth]{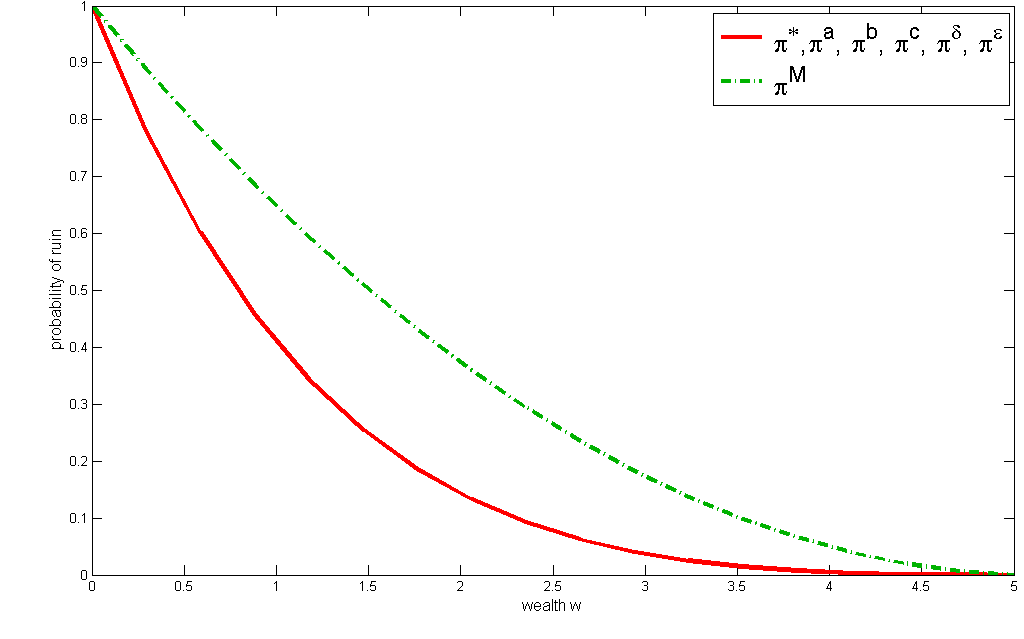}
    \caption{Performance of the investment strategies described in Observation 3 (of Section 6).  Speed of mean reversion=$0.2$. Correlation $\rho=0.5$. Initial volatility $\sigma_0 = \sigma_m=0.25$.}
  \label{fig:diffStrat109b}
\end{figure}

\begin{figure}
  \centering
   \includegraphics[width=0.9\textwidth]{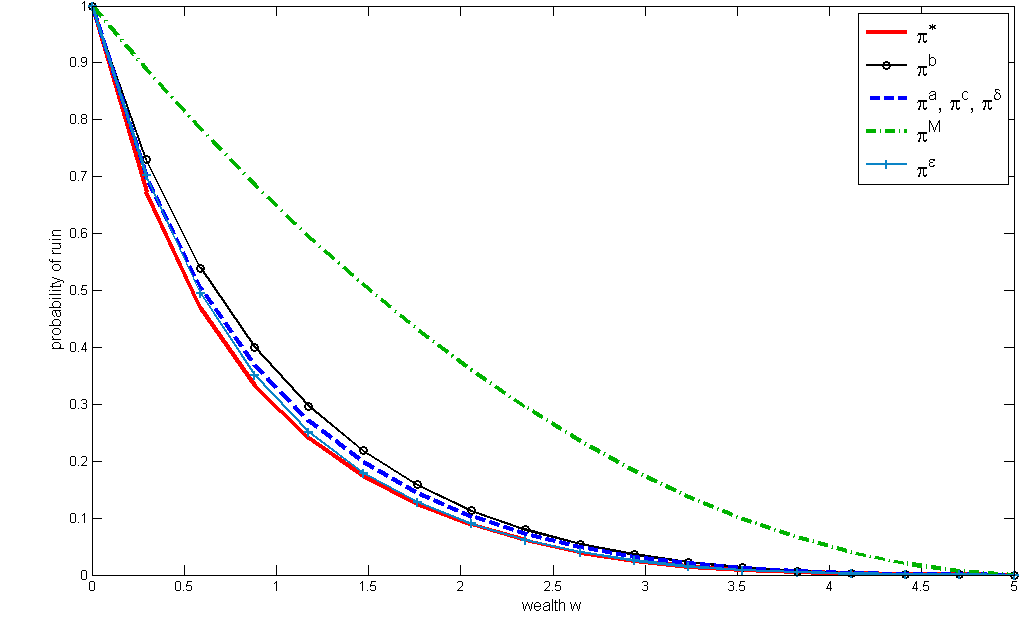} 
     \caption{Performance of the investment strategies described in Observation 3 (of Section 6).  Speed of mean reversion=$0.2$. Correlation $\rho=0.5$. Initial volatility $\sigma_0=0.1$. }
  \label{fig:diffStrat109c}
\end{figure}

In Figure~\ref{fig:diffStrat109}-\ref{fig:diffStrat109c} we observe that the performance of $\pi^c$ and $\pi^{\eps}$ are almost as good as the optimal strategy $\pi^*$. (Here we are considering a medium mean reversion speed. When the mean reversion speed is much smaller, then $\pi^{\delta}$ would be a better investment strategy.) Moreover, their performances are robust, in that, they do not depend on the initial volatility $\sigma_0$. This should be contrasted to $\pi^a$ and $\pi^b$. The former performs relatively well when $\sigma_0$ is small, whereas the latter performs better when $\sigma_0$ is large. When $\sigma_0=\sigma_m$, all strategies perform as well as the optimal strategy. Also, observe that for wealthy or very poor individuals the choice of the strategy does not matter as long as they invest in the stock market. The difference is for the individuals who lie in between. 

As a result, we conclude that if the individual wants to minimize her probability of ruin in a stochastic volatility environment, she can still use the investment that is optimal for the constant volatility environment. She simply needs to update the volatility in that formula whenever the volatility changes significantly. 

\bibliographystyle{model2-names}
\bibliography{BHYrefs}

\newpage

\end{document}